\documentclass[11pt]{article}
\usepackage{graphicx}
\usepackage{amsmath,amssymb,amsthm,amsfonts}
\usepackage{color}
\usepackage{amssymb}
\usepackage{cite}
\newtheorem{thm}{Theorem}[section]

\newtheorem{lem}[thm]{Lemma}

\theoremstyle{definition}
\newtheorem{defn}{Definition}[section]
\theoremstyle{remark}
\usepackage{appendix}
\newtheorem{rem}{Remark}[section]

\numberwithin{equation}{section}

\DeclareMathSymbol{\C}{\mathalpha}{AMSb}{"43}

\newcommand{\alp}{\alpha}

\newcommand{\R}{{\mathbb{R}}}

\def\R{{\mathbb R}}
\def\C{{\mathbb C}}

\allowdisplaybreaks[4]

\newcommand{\bsub}{\begin{subequations}}
\newcommand{\esub}{\end{subequations}$\!$}

\begin{document}
	
\title{Ground States of Fermionic Nonlinear Schr\"{o}dinger Systems with Coulomb Potential I: The $L^2$-Subcritical Case}
\date{}
	
\author{Bin Chen\thanks{Email:  binchenmath@mails.ccnu.edu.cn.} \; and\,  Yujin Guo\thanks{Email: yguo@math.ccnu.edu.cn. Y. Guo is partially supported by NSFC under Grants 12225106 and 11931012.}\\
\small \it	School of Mathematics and Statistics,\\
\small \it  Key Laboratory of Nonlinear Analysis $\&$ Applications (Ministry of Education),\\
\small \it Central China Normal University, Wuhan 430079, P. R. China\\}
\date{\today}

\smallbreak \maketitle
\begin{abstract}
We consider ground states of the $N$ coupled fermionic nonlinear Schr\"{o}dinger systems with the Coulomb potential $V(x)$ in the $L^2$-subcritical case. By studying the associated constraint variational problem, we prove the existence of ground states for the system with any parameter $\alpha>0$, which represents the attractive strength of the non-relativistic quantum particles. The limiting behavior of ground states for the system is also analyzed as $\alpha\to\infty$, where the mass concentrates at one of the singular points for the Coulomb potential $V(x)$.
\end{abstract}
	
\vskip 0.05truein


	
\vskip 0.2truein

\section{Introduction}
The quantum many-body problem has received a lot of attentions since it was proposed as a precise mathematical form in 1926 (cf. \cite{1926}). A system of $N$ (spinless) non-relativistic particles in quantum mechanics can be described by an energy functional $\Psi\mapsto\mathcal{E}(\Psi)$, see \cite{ii,i,geomrtric, 1987, 1977, L3}, where $\Psi\in H^1(\R^{3N}, \C)$ is a normalized wave function. In this paper, we study ground states $ (u_1, \cdots, u_N)\in \big(H^1(\R^3,\R)\big)^N$ of the following fermionic nonlinear Schr\"{o}dinger (NLS) system
\begin{align}\label{j1}
\arraycolsep=1.5pt
\left\{\begin{array}{lll}
	\Big[-\Delta+V(x)-\alpha^{2p-2}\Big(\sum_{j=1}^N|u_j|^2\Big)^{p-1}\Big]u_i=\mu_i u_i\ \,
	&\mathrm{in}\ \ \R^3,\ \ \alpha>0,\\[3mm]
	(u_i,u_j)_{L^2(\R^3,\R)}=\delta_{ij},\ \ i,\ j=1,\cdots,N\in\mathbb{N}^+,
	\end{array}\right.
\end{align}
where $ 1<p<\frac{5}{3}$, and the function $V(x)$ is an attractive Coulomb potential of the form
\begin{equation}\label{1:1A}
V(x)=-\sum _{k=1}^K|x-y_k|^{-1} \ \ \, \mbox{in}\ \, \R^3,\ \ y_k\neq y_l \,\ \text{for}\,\   k\neq l .
\end{equation}
The NLS system \eqref{j1} arises (cf. \cite{geomrtric}) from the following energy functional of $N$ spinless non-relativistic quantum particles:
\begin{equation}\label{1.3}
\mathcal{E}_\alpha(\Psi):=\sum_{i=1}^N\int_{\R^{3N}}\Big(|\nabla_{x_i}\Psi|^2+V(x_i)|\Psi|^2\Big)dx_1\cdots dx_N-\frac{\alpha^{2p-2}}{p}\int_{\R^3}\rho_\Psi^{p}dx,\ \,\alpha>0,
\end{equation}
where $1<p<\frac{5}{3}$, $\Psi\in H^1(\R^{3N}, \C)$, and the one-particle density $\rho_\Psi$ associated to $\Psi$ is defined as
\begin{equation*}
\begin{split}
\rho_\Psi(x):=&\int_{\R^{3(N-1)}}|\Psi(x,x_2,\cdots,x_N)|^2dx_2\cdots dx_N\\
&+\cdots+\int_{\R^{3(N-1)}}|\Psi(x_1,\cdots,x_{N-1},x)|^2dx_1\cdots dx_{N-1}.
\end{split}
\end{equation*}
The above parameter $\alpha>0$ represents the attractive strength of the non-relativistic quantum particles, and the attractive Coulomb potential $V(x)$ of \eqref{1:1A} is usually generated by a molecule (see for example \cite{1974, 1977, 1987, L3, geomrtric}).  We refer the reader to   \cite{ii,i, geomrtric} and the references therein for more physical motivations of \eqref{1.3}.

It is known that all elementary particles in nature are divided mainly into two classes, in terms of the spin quantum numbers, which are called bosons and fermions. Specially, if the above system \eqref{1.3} contains only $N$ identical bosons or fermions,  
then the corresponding bosonic or fermionic constraint variational problem satisfies
\begin{equation}\label{1.5}
E_{b/f}(N):=\inf\Big\{\mathcal{E}_\alpha(\Psi):\Psi\ \mathrm{is \ bosonic\ or\ fermionic},\ \|\Psi\|^2_2=1,\ \Psi\in H^1(\R^{3N},\C)\Big\}.
\end{equation}
For convenience, we denote $\vee^NL^2(\R^3,\C)\  \big(\text{resp}.\  \wedge^N L^2(\R^3,\C)\big)$ the subspace of $L^2(\R^{3N},\C)$ consisting of all symmetric (resp. antisymmetric) wave functions. 

For bosons, which satisfy Bose-Einstein statistics, the corresponding wave function $\Psi$ is symmetric, i.e., $\Psi\in \vee^NL^2(\R^3,\C)$ (see {\cite[Section 3]{L3}}). Taking $u\in L^2(\R^3,\C)$ with $\|u\|_2=1$ and letting $\Psi:=\Pi_{i=1}^Nu(x_i)$, one can get that $\Psi\in \vee^NL^2(\R^3,\C)$, $\|\Psi\|_2=1$ and $\rho_\Psi=N|u|^2$. Therefore, as commented in \cite[Remark 8]{i}, the infimum $E_{b}(N)$ of (\ref{1.5}) can be then reduced equivalently to the following form
\begin{align*}
E_{b}(N)&=NI(a)\nonumber\\
:&= N\inf\Big\{\int_{\R^3}\big(|\nabla u|^2+V(x)u^2\big)dx
-\frac{a^{2p-2}}{p}\int_{\R^3}u^{2p}dx:\ \|u\|_2^2=1,\ u\in H^1(\R^{3},\R)\Big\},
\end{align*}
where  the potential $V(x)$ is as in (\ref{1:1A}), and $a:=\alpha N^{1/2}>0$. When $1<p<\frac{5}{3}$,  the constraint variational problem $I(a)$ is usually referred to as an  {\em $L^2$-subcritical} problem (see \cite{Cazenave, l2}),  which has attracted a lot of attentions since 1970s, see for example \cite{Cazenave,concen,concen2,m,Rabinowitz} and the references therein. More precisely, the authors in \cite{concen,concen2,Rabinowitz} proved the existence of minimizers for $I(a)$. Furthermore,  the limiting behavior, the local uniqueness and some other analytical properties of minimizers for $I(a)$ were also investigated in \cite{Cazenave,m} and the references therein.

For fermions, which satisfy Fermi-Dirac  statistics, the corresponding wave function  $\Psi$  is antisymmetric, i.e., $\Psi\in\wedge^NL^2(\R^3,\C)$. By the Pauli exclusion principle, the simplest example of  antisymmetric functions is that $\Psi$ is a Slater determinant, i.e., $\Psi=(N!)^{-1/2}\mathrm{det}\{u_i(x_j)\}_{i,j=1}^N$, where $u_i\in L^2(\R^3,\C)$ and $(u_i, u_j)_{L^2}=\delta_{ij},\ i,\ j=1,\cdots, N$. In this case,
we have $\|\Psi\|_{2}=1$ and the energy functional $\mathcal{E}_\alpha(\Psi)$  of (\ref{1.3}) becomes
\begin{equation}\label{1.0a}
	\mathcal{E}_\alpha(\Psi)=\mathcal{E}_\alpha(u_1,\cdots,u_N),
\end{equation}
where $\mathcal{E}_\alpha(u_1, \cdots, u_N)$ is defined by
\begin{equation}\label{functional}
\begin{split}
\mathcal{E}_\alpha(u_1, \cdots, u_N):=&\sum_{i=1}^N\int_{\R^3}\big(|\nabla u_i|^2+V(x)| u_i|^2\big)dx-\frac{\alpha^{2p-2}}{p}\int_{\R^3}\Big(\sum_{i=1}^N| u_i|^2\Big)^pdx,
\end{split}
\end{equation}
and the potential $V(x)$ is as in (\ref{1:1A}), together with $1<p<\frac{5}{3}$ and $\alpha>0$. Applying (\ref{1.5}) and (\ref{1.0a}), we shall illustrate in Appendix A that
\begin{equation}\label{1.0}
\begin{split}
E_f(N)&=J_\alpha(N)\\
:&=\inf\Big\{\mathcal{E}_\alpha(u_1,\cdots,u_N):\ u_1,\cdots,u_N\in H^1(\R^{3},\R),\ (u_i,u_j)_{L^2}=\delta_{ij}\Big\},
\end{split}
\end{equation}
where  the energy functional $\mathcal{E}_\alpha(u_1, \cdots, u_N)$ is given by (\ref{functional}).


In view of the above facts, in the present paper we focus on the minimization problem $J_{\alpha}(N)$ defined in (\ref{1.0}) with $1<p<\frac{5}{3}$. Stimulated by the many-body boson problems mentioned as above, we refer to this situation as the {\em $L^2$-subcritical case} of $J_\alpha(N)$.
The {\em$L^2$-critical case} (i.e., $p=\frac{5}{3}$) of $J_\alpha(N)$  is however left to the companion work \cite{critical}. As for the {\em$L^2$-supercritical case} (i.e., $p>\frac{5}{3}$) of $J_\alpha(N)$, a standard scaling argument gives that $J_\alpha(N)=-\infty$ for any $\alpha>0$ and $N\in\mathbb{N}^+$, which  thus yields that $J_\alpha(N)$ does not admit any minimizer for any $\alpha>0$ and $N\in\mathbb{N}^+$.  The main purpose of the present paper is to address the limiting behavior of minimizers for the system $J_{\alpha}(N)$ as $\alpha\to\infty$, where $1<p<\frac{5}{3}$. As far as we know, this seems the first work on the  asymptotics of  the $N$ coupled  {\em fermionic}  nonlinear Schr\"{o}dinger systems.

We now introduce the concept of ground states for the system \eqref{j1}.

\begin{defn}\label{dfn:1} (\emph{Ground states}). A system $(u_1,\cdots, u_N )\in \big(H^1(\R^{3},\R )\big)^N$ with $(u_i, u_j)_{L^2}$ $=\delta_{ij}$ is called a ground state of \eqref{j1}, if  it solves the system \eqref{j1}, where $\mu_1<\mu_2\leq\cdots\leq\mu_N\leq0$ are the $N$ first eigenvalues (counted with multiplicity) of the operator
\begin{equation}\label{1.12M}
H_V:=-\Delta+V(x)-\alpha^{2p-2}\Big(\sum_{j=1}^Nu_j^2\Big)^{p-1}\ \  \mbox{in} \ \, \R^3.
\end{equation}
\end{defn}

\noindent
The first result of the present paper is concerned with the following existence of minimizers for  $J_\alpha(N)$ defined in \eqref{1.0}.


\begin{thm}\label{th1}
For any $\alpha>0$, $N\in\mathbb{N}^+$ and  $p\in\big(1,\frac{5}{3}\big)$, the problem $J_\alpha(N)$ defined in \eqref{1.0} has at least one minimizer $(u_1^\alpha,\cdots,u_N^\alpha)$, which is a ground state of  the following  system:
\begin{equation}\label{thmA:1}
\Big(-\Delta+V(x)-\alpha^{2p-2}\Big(\sum_{j=1}^N|u_j^\alpha|^2\Big)^{p-1}\Big)u_i^\alpha=\mu^\alpha_i u^\alpha_i \ \ in\, \ \R^3,\ \ i=1,\cdots,N.
\end{equation}
Here  $(u_i^\alpha,u_j^\alpha)=\delta_{ij}$, and $\mu_1^\alpha<\mu_2^\alpha\leq\cdots\leq \mu_N^\alpha<0$ are the  $N$ first  eigenvalues, counted with multiplicity, of the Schr\"{o}dinger operator $H_V$ defined in (\ref{1.12M}).
\end{thm}

The proof of Theorem \ref{th1} is based on an adaptation of the classical concentration compactness principle (cf. \cite[Sect. 3.3]{begain}), for which  reason we shall establish in Lemma \ref{lem2.4} a strict binding inequality.  Theorem \ref{th1} shows that for  $N\in\mathbb{N}^+$ and  $p\in\big(1,\frac{5}{3}\big)$,  $J_\alpha(N)$ admits at least one minimizer for all $\alpha>0$, which is a ground state of the fermionic NLS system \eqref{j1}. Moreover, the existence of Theorem \ref{th1} can be extended naturally to the general dimensional case $\R^d$ with $d\geq3$,   and to more general  potentials $V(x)$.
For simplicity we however do not pursue these general situations.

Denote by  $J_\alpha^\infty(N)$ the variational problem $J_\alpha(N)$ without the potential $V(x)$:
\begin{equation} \label{jinfty}
\begin{split}
	J_\alpha^\infty(N):=\inf\Big\{&\mathcal{E}_\alpha^\infty(u_1,\cdots,u_N):u_1,\cdots,u_N\in H^1(\R^{3},\R),\  \\
&(u_i,u_j)_{L^2}=\delta_{ij}, \ i,j=1,\cdots,N\Big\},\ \alpha>0,\ N\in\mathbb{N}^+,
\end{split}
\end{equation}
where the energy functional $\mathcal{E}_\alpha^\infty(u_1,\cdots,u_N)$ satisfies
\begin{gather*}
\mathcal{E}_\alpha^\infty(u_1,\cdots,u_N):=\sum_{i=1}^N\int_{\R^3}|\nabla u_i|^2dx-\frac{\alpha^{2p-2}}{p}\int_{\R^3}\Big(\sum_{i=1}^N| u_i|^2\Big)^pdx, \,\ p\in\big(1,\, \frac{5}{3}\big).
\end{gather*}
One can check that
\begin{align}\label{1.14}
\mathcal{E}_\alpha^\infty(u^\alpha_1,\cdots,u^\alpha_N)=\alpha^{\frac{4(p-1)}{2-3(p-1)}}\mathcal{E}_1^\infty(u_1,\cdots,u_N)\ \ \ \text{and}\ \ \ J_\alpha^\infty(N)=\alpha^{\frac{4(p-1)}{2-3(p-1)}}J_1^\infty(N),
\end{align}
where $u_i^\alpha(x)\equiv\alpha^{\frac{3(p-1)}{2-3(p-1)}}u_i\big(\alpha^{\frac{2(p-1)}{2-3(p-1)}}x\big)$ in $\R^3$ for $i=1, \cdots, N$.
We remark that the existence of minimizers for $J_1^\infty(N)$ was addressed  in \cite[Theorem 3]{i} by applying \cite[Theorem 27]{geomrtric}, where the authors however obtained the compactness of the minimizing sequences instead by the geometric methods of nonlinear many-body quantum systems.  Following \cite[Theorem 4]{i}, there exists a constant $p_c\in(1,\frac{5}{3}]$ such that for any $p\in(1, p_c)$ and $N\in\mathbb{N}^+$,  $J_1^\infty(N)$  admits at least one minimizer. This further  yields from \eqref{1.14} that for any $p\in(1, p_c)$ and $N\in\mathbb{N}^+$,  $J_\alpha^\infty(N)$ possesses minimizers for all $\alpha>0$. Different from  \cite[Theorem 3]{i}, we emphasize that the existence of  Theorem \ref{th1} is proved in the whole $L^2$-subcritical range of $p$, i.e., $p\in(1, 5/3)$. However, our proof of Theorem \ref{th1} is more involved than that of \cite[Theorem 3]{i},  due to the appearance of the Coulomb potential $V(x)$. 

Let $p_c\in(1,\frac{5}{3}]$ be given as stated above, and we next focus on the limiting behavior of minimizers for $J_\alpha(N)$ as $\alpha\rightarrow\infty$, where $p\in(1,p_c)$ and $N\in\mathbb{N}^+$. The main result of the present paper can be then stated as the following theorem.

\begin{thm}\label{th2} Let $(u_1^\alpha,\cdots,u_N^\alpha)$ be a minimizer of $J_\alpha(N)$ defined in \eqref{1.0} for $p\in(1,p_c)$, which is a ground state of \eqref{thmA:1}. Then for any sequence $\{\alpha_n\}$ satisfying $\alpha_n\rightarrow\infty$ as $n\rightarrow\infty$, there exists a subsequence, still denoted by $\{\alpha_n\}$, of $\{\alpha_n\}$ such that
\begin{equation}\label{1.16}
\begin{split}
\hat{w}_i^{\alpha_n}(x):&= \alpha_n^{\frac{-3(p-1)}{2-3(p-1)}}u_i^{\alpha_n}
\big(\alpha_n^{\frac{-2(p-1)}{2-3(p-1)}}x+z_n\big)\\[1mm]
&\to \hat{w}_i(x)\ \,   strongly\ in \, \ L^\infty(\R^3)\, \ as\, \ n\rightarrow\infty,\ \, i=1, \cdots, N,
\end{split}
\end{equation}
where  $(\hat{w}_1,\cdots, \hat{w}_N)$ is a minimizer of $J_1^\infty(N)$  given by \eqref{jinfty}, and $z_n\in \R^3$ is a global maximal point of $\sum_{i=1}^N|u_i^{\alpha_n}|^2$ satisfying
\begin{equation}\label{1.A4}
|z_n-y_k|\leq C\alpha_n^{\frac{-2(p-1)}{2-3(p-1)}}\ \ as\ \ n\to\infty
\end{equation}
for some $y_k\in\{y_1,\cdots,y_K\}$ given in \eqref{1:1A}. Moreover, there exist constants $\theta>0$ and $C(\theta)>0$, independent of $n>0$, such that
\begin{equation}\label{1.4}
\sum_{i=1}^N|\hat{w}_i^{\alpha_n}(x)|^2\leq C(\theta)e^{-\theta|x|}\  \  uniformly\, \ in\ \, \R^3\, \  as \, \ n\to\infty.
\end{equation}
\end{thm}

The proof of Theorem \ref{th2} follows from a detailed analysis of the global minimum energy $J_\alpha(N)$ and the associated fermionic system (\ref{thmA:1}) as $\alpha\rightarrow\infty$. We thus make full use of the following Gagliardo-Nirenberg-Sobolev inequality for the orthonormal system: for any $(u_1,\cdots,u_N)\in \big(H^1(\R^3)\big)^N$ with $(u_i,u_j)_{L^2}=\delta_{ij}$, $i,j=1,\cdots,N$,
\begin{eqnarray}\label{gn}
\sum_{i=1}^{N}\int_{\R^3}|\nabla u_i|^2dx\geq K(p,N)\Big(\int_{\R^3}\Big(\sum_{i=1}^{N}|u_i|^2\Big)^pdx\Big)^{\frac{2}{3(p-1)}},\ \, 1<p<\frac{5}{3},
\end{eqnarray}
where the constant $K(p,N)>0$ satisfies
$$
K(p,N):=(p-1)|J_1^\infty(N)|^{-\frac{2-3(p-1)}{3(p-1)}}\Big(\frac{3}{2p}\Big)^{\frac{2}{3(p-1)}}
\Big(\frac{5}{3}-p\Big)^{\frac{2-3(p-1)}{3(p-1)}}>0,
$$
and the identity of \eqref{gn} is achieved
at a minimizer $(\hat{w}_1,\cdots,\hat{w}_N)$ of $J_1^\infty(N)$ defined in (\ref{jinfty}), see \cite{i} for more details.
On the other hand, the $L^\infty$-uniform convergence of Theorem \ref{th2} shows that the minimizer of $J_\alpha(N)$ blows up near some singular point $y_k$, i.e., a global minimum point, of the Coulomb potential $V(x)=-\Sigma_{k=1}^K|x-y_k|^{-1}$ as $\alpha\rightarrow\infty$.  It is thus interesting to further investigate the exact point $y_k$ among $\{y_1,\cdots, y_K\}$.

The $L^\infty$-uniform convergence of \eqref{1.16} depends strongly on the uniformly exponential decay of \eqref{1.4} in $n>0$, which cannot be however established by the standard comparison principle, due to the singularities of the Coulomb potential $V(x)$. Actually,
it follows from \eqref{thmA:1} that for $i=1,\cdots,N$,  the function $\hat{w}_i^{\alpha_n}$ defined in (\ref{1.16}) solves
\begin{equation}\label{1:120}
-\Delta \hat{w}_i^{\alpha_n}+\epsilon_{\alpha_n}^2V(\epsilon_{\alpha_n}\cdot+z_n)\hat{w}_i^{\alpha_n}
-\Big(\sum_{i=1}^N|\hat{w}_i^{\alpha_n}|^2\Big)^{p-1}\hat{w}_i^{\alpha_n}=\epsilon_{\alpha_n}^2\mu_i^{\alpha_n} \hat{w}_i^{\alpha_n}\ \ \ \mathrm{in}\ \, \R^3,
\end{equation}
where $\epsilon_{\alpha_n}:=\alpha_n^{\frac{-2(p-1)}{2-3(p-1)}}\to0$ as $n\to\infty$, and the Coulomb potential term  satisfies
$$\epsilon_{\alpha_n}^2V(\epsilon_{\alpha_n} x+z_n)=-\epsilon_{\alpha_n}\sum_{k=1}^{K}\Big| x-\frac{y_k-z_n}{\epsilon_{\alpha_n}}\Big|^{-1}\,.$$
It unfortunately yields from \eqref{1.A4} that the Coulomb potential term of (\ref{1:120}) is  singular for sufficiently large $|x|$  as $n\to\infty$, and hence the standard comparison principle is not applicable for (\ref{1:120}).
To overcome this difficulty,  we shall prove in Lemma \ref{lem:3.3} the uniformly exponential decay of \eqref{1.4}, by employing the Green's function to analyze the elliptic problem (\ref{1:120}).


This paper is organized as follows. In Section 2,  we shall address the proof of Theorem \ref{th1} on the existence of minimizers for $J_\alpha(N)$.  Section 3 is devoted to the proof of Theorem \ref{th2} on the mass concentration of minimizers for $J_\alpha(N)$. The relation (\ref{1.0}) and Lemma \ref{lem2.4} are finally proved in Appendix A for the reader's convenience.

\section{Existence of Minimizers for $J_\alpha(N)$}

The main purpose of this section is to establish Theorem \ref{th1} on the existence of minimizers for  $J_\alpha(N)$, where $\alpha>0$ and $N\in\mathbb{N}^+$ are arbitrary. We shall first establish several lemmas, based on which Theorem \ref{th1}  is finally proved in Subsection 2.1.

We start by introducing the following minimization problem
\begin{equation}\label{1.2}
E_\alpha(\lambda):=\inf\limits_{\gamma\in\mathcal{K}_\lambda} \mathcal{E}_\alpha(\gamma),\ \ \lambda >0, \ \ \alpha >0,
\end{equation}
where
\begin{equation}\label{2:2}
	\mathcal{E}_\alpha(\gamma):=\mathrm{Tr}\big(-\Delta+V(x)\big) \gamma-\frac{\alpha^{2p-2}}{p}\int_{\R^3}\rho_\gamma^pdx,\ \ \ 1<p<\frac{5}{3},
\end{equation}
\begin{equation}\label{1.11a}
	\mathcal{K}_{\lambda}:=\big\{\gamma\in \mathcal{B}\big(L^2(\R^3, \R)\big):\ 0\leq\gamma=\gamma^*\leq 1,\ \mathrm{Tr}\gamma=\lambda,\ \mathrm{Tr}(-\Delta \gamma)<\infty\big\},
\end{equation}
and $\mathcal{B}\big(L^2(\R^3)\big)$ denotes the set of bounded linear operators on $L^2(\R^3)$.
In \eqref{2:2},  the potential $V(x)\le 0$ is as in (\ref{1:1A}), and the function $\rho_\gamma(x)$  is defined below by \eqref{2.2E}. 
The advantage of  (\ref{1.2}) lies in the fact that $\mathcal{K}_{\lambda}$ is convex.
Note from the spectral theorem (see \cite{i} and the references therein) that  for any $\gamma\in\mathcal{K}_\lambda$, there exist an orthonormal basis $\{u_i\}$ of $L^2(\R^3)$ and a sequence $\{n_i\}\subset\R$ such that the operator $\gamma $ satisfies
\begin{equation}\label{2.2D}
\gamma=\sum_{i\ge 1}n_i|u_i\rangle\langle u_i|,
\end{equation}
where $0\leq n_i\leq1$,  $\sum_{i\ge 1}n_i=\lambda$ and
\begin{equation}\label{2.2DD}
\gamma\varphi(x)=\sum_{i\ge 1}n_iu_i(x)(u_i, \varphi) \ \ \text{for any}\ \ \varphi\in L^2(\R^3).
\end{equation}
Associated to the operator $\gamma$, the function $\rho_\gamma(x)$ in (\ref{2:2})
 is defined as
\begin{equation}\label{2.2E}
\rho_\gamma(x):=\gamma(x,x),
\end{equation}
where $\gamma(x,y)=\sum_{i\ge 1}n_iu_i(x)u_i(y) $ denotes the integral kernel of the operator $\gamma $.
By denoting  $P_j=-i\partial _j$, we then have
\begin{equation}\label{2.2EA}
\mathrm{Tr}(-\Delta \gamma):=\sum^{3}_{j=1}\mathrm{Tr}(P_j\gamma P_j)=\sum _{i\ge 1}n_i\int_{\R^3}|\nabla u_i(x)|^2dx.
\end{equation}

We next note from \cite{L3,L2} the following Lieb-Thirring inequality:
\begin{equation}\label{lt}
	\|\gamma\|^\frac{2}{3}\mathrm{Tr}(-\Delta \gamma)\geq c_{\mathrm{LT}}\int_{\R^3}\rho_{\gamma}^{\frac{5}{3}}dx, \ \ \forall\ \gamma\in\mathcal{K}_\lambda\ \ \mathrm{and}\ \lambda>0,
\end{equation}
where $0\le \|\gamma\|=\max(n_i)\le 1$ denotes the norm of the operator $\gamma$ on $L^2(\R^3)$, and the constant $c_{\mathrm{LT}}>0$ is independent of $\lambda$. We also recall  (cf. \cite{ho}) the following Hoffmann-Ostenhof inequality:
\begin{equation}\label{ho}
	\mathrm{Tr}(-\Delta \gamma)\geq\int_{\R^3}|\nabla\sqrt{\rho_\gamma}|^2dx, \ \ \forall\ \gamma\in\mathcal{K}_\lambda\ \ \mathrm{and}\ \lambda>0.
\end{equation}
Applying (\ref{2.2D})--(\ref{ho}), we have the following equivalence.

\begin{lem}\label{lem:2.1}  Suppose the problem $E_\alpha(\lambda)$ is defined by (\ref{1.2}), where $p\in(1, \frac{5}{3})$ and  $ \alpha>0$. Then we have
\begin{align}\label{lem1}
E_\alpha(\lambda)=&\inf\Big\{  \mathcal{E}_\alpha(\gamma):\ \gamma=\sum_{i=1}^{N-1}|u_i\rangle\langle u_i|+(\lambda-N+1)|u_N\rangle\langle u_N|,\Big.\nonumber\\
&\ \ \ \ \ \ \ \Big.u_i\in H^1(\R^3)\ \mbox{with}\ (u_i,u_j)=\delta_{ij},\  i,j=1,\cdots,N\Big\}, \ \forall \ \lambda>0,
\end{align}
where the functional $\mathcal{E}_\alpha(\gamma)$ is as in (\ref{2:2}), and $N$ is the smallest integer such that $\lambda\leq N$.
\end{lem}

\begin{rem}\label{rem2:2}
Following the argument of (\ref{2.2D})--(\ref{2.2EA}) and the definition of Trace, one can obtain from (\ref{1.2})  that for any $\alpha >0$ and $N\in \mathbb{N}^+$,
\begin{align*}\label{rem2:2A}
 J_\alpha(N)
 =&\inf\Big\{\mathcal{E}_\alpha(u_1, \cdots, u_N): u_i\in H^1(\R^3)\ \mbox{with}\  (u_i,u_j)=\delta_{ij}\Big\}\nonumber\\
 =&\inf\Big\{  \mathcal{E}_\alpha(\gamma):\ \gamma=\sum_{i=1}^{N}|u_i\rangle\langle u_i|,\ u_i\in H^1(\R^3)\ \mbox{with}\ (u_i,u_j)=\delta_{ij}\Big\}\\
=&E_\alpha(N),\nonumber
\end{align*}
where $J_\alpha(N)$ is defined in \eqref{1.0}, and the last identity follows from Lemma \ref{lem:2.1}. Therefore,  the definition of the problem $E_\alpha(N)$ in (\ref{1.2}) is essentially consistent with the problem $J_\alpha(N)$.
\end{rem}

Since the proof of Lemma \ref{lem:2.1} is similar to that of \cite[Lemma 11]{i}, we omit the detailed proof for simplicity. Associated to the minimization problem $E_\alpha(\lambda)$, we now define the minimization problem without the external potential $V(x)$:
\begin{equation}\label{D:1.2}
E^\infty_\alpha(\lambda):=\inf\limits_{\gamma\in\mathcal{K}_\lambda} \mathcal{E}^\infty_\alpha(\gamma),\ \ \lambda >0, \ \ \alpha >0,
\end{equation}
where the constraint $\mathcal{K}_\lambda$ is as in (\ref{1.11a}), and
\begin{equation*}\label{D:2:2}
	\mathcal{E}^\infty_\alpha(\gamma):=\mathrm{Tr}(-\Delta \gamma)-\frac{\alpha^{2p-2}}{p}\int_{\R^3}\rho_\gamma^pdx.
\end{equation*}
The following lemma presents some basic properties of the problem $E_\alpha(\lambda): \R^+\mapsto \R $, which are crucial  for the proof of Theorem \ref{th1}.

\begin{lem}\label{lem:2.2}
Suppose the problem $E_\alpha(\lambda)$ is defined by (\ref{1.2}), where $p\in(1, \frac{5}{3})$ and  $ \alpha>0$. Then we have the following assertions:
\begin{enumerate}
\item The energy estimate $-\infty<E_\alpha(\lambda)<0$ holds for any $\lambda>0$.

\item It holds that
$$
E_\alpha(\lambda+\lambda')\leq E_\alpha(\lambda)+E^\infty_\alpha(\lambda'), \ \ \forall \ \lambda',\ \lambda>0,
$$
where $E^\infty_\alpha(\cdot)$ is defined by (\ref{D:1.2}).

\item  $ E_\alpha(\lambda)$ decreases strictly and is Lipschitz continuous in $\lambda>0$.

\item  $E_\alpha(\lambda)$ is concave on each interval $(N-1, N)$ for all integer $N\in \mathbb{N}^+$.
\end{enumerate}
\end{lem}

Applying  Hardy's inequality, one can get that
\begin{equation}\label{2:hardy}
|x|^{-1}\leq \varepsilon(-\Delta)+4\varepsilon^{-1}\ \ \  \mbox{for any}\ \varepsilon>0.
\end{equation}
Following the inequality (\ref{2:hardy}) and the Lieb-Thirring inequality \eqref{lt}, one can further obtain that $E_\alpha(\lambda)>-\infty$ holds for any $\lambda>0$. Because the rest parts of Lemma \ref{lem:2.2} can be proved in a similar way of \cite[Lemma 12]{i}, we leave the detailed proof of Lemma \ref{lem:2.2} to the interested reader. Applying the above two lemmas, we next address the following properties of minimizers for $E_\alpha(\lambda)$.

\begin{lem}\label{lem2.3}
Suppose the problem $E_\alpha(\lambda)$ is defined by (\ref{1.2}) for $\lambda>0$, where $p\in(1, \frac{5}{3})$ and  $ \alpha>0$. Then we have
\begin{enumerate}
\item  If  $E_\alpha(\lambda)$ possesses minimizers, then one of them must be of the form
\begin{equation}\label{2.10}
\gamma:=\sum_{i=1}^{N-1}|u_i\rangle\langle u_i|+(\lambda-N+1)|u_N\rangle\langle u_N|,\ \ \ (u_i, u_j)=\delta_{ij},
\end{equation}
where $N$ is the smallest integer such that $\lambda\leq N$, and the orthonormal family $(u_i, \cdots,u_N)$ satisfies
\begin{equation}\label{6}
\left(-\Delta+V(x)-\alpha^{2p-2}\rho_{\gamma}^{p-1}\right)u_i=\mu_i u_i\ \ in\ \,\R^3,\ \ i=1,\cdots,N.
\end{equation}
Here $\rho_{\gamma}=\sum_{i=1}^{N-1}u_i^2+(\lambda-N+1)u_N^2$,  $\mu_i$ are the $N$ first eigenvalues, counted with multiplicity, of the operator $H_\gamma:=-\Delta+V(x)-\alpha^{2p-2}\rho_{\gamma}^{p-1}$ in $\R^3$, and satisfy $\mu_1<\mu_2\leq\cdots\leq\mu_N<0$.

\item   Let  $\gamma$ be a minimizer of $E_\alpha(\lambda)$ in the form of \eqref{2.10}, then the following estimates hold:
\begin{gather}\label{11}
	C^{-1}(1+|x|)^{-1}e^{-\sqrt{|\mu_1|}|x|}\leq u_1(x)\leq C(1+|x|)^{\frac{K}{\sqrt{|\mu_1|}}-1}e^{-\sqrt{|\mu_1|}|x|}\ \ in\ \,\R^3,\\[3mm]
	|u_i(x)|\leq C(1+|x|)^{\frac{K}{\sqrt{|\mu_i|}}-1}e^{-\sqrt{|\mu_i|}|x|} \ \ in\ \, \R^3,\ \ i=2,\cdots, N,\label{2.7}
\end{gather}
where the constant $K>0$ is as in (\ref{1:1A}), and the constant $C>0$ depends on $\alpha>0$ and $\|\rho_{\gamma}\|_{L^{3}(\R^3)}$.
\end{enumerate}
\end{lem}



\noindent \textbf{Proof.} 1. Let $\gamma$ be a minimizer of $E_\alpha(\lambda)$.  We first claim that $\gamma$ is an optimizer of the infimum
\begin{eqnarray}\label{2.8}
	\inf\limits_{\gamma'\in\mathcal{K}_{\lambda}}\mathrm{Tr}H_\gamma(\gamma'),\ \ \ \ \mathrm{where}\ \  H_{\gamma}:=-\Delta+V(x)-\alpha^{2p-2}\rho_{\gamma}^{p-1}\ \ \mathrm{in}\ \,\R^3,
\end{eqnarray}
and any optimizer of  \eqref{2.8} is also a minimizer for $E_\alpha(\lambda)$. Indeed, since $p>1$, it holds for any $\gamma'\in\mathcal{K}_{\lambda}$,
\begin{equation}\label{5}
\begin{split}
	\mathcal{E}_\alpha(\gamma')=&\ \mathcal{E}_\alpha(\gamma)
	 +\mathrm{Tr}H_\gamma(\gamma'-\gamma)-\frac{\alpha^{2p-2}}{p}\int_{\R^3}\Big[\rho_{\gamma'}^{p}-\rho_\gamma^{p}-p\rho_\gamma^{p-1}
	(\rho_{\gamma'}-\rho_\gamma)\Big]dx \\
	\leq& \
	\mathcal{E}_\alpha(\gamma)
	+\mathrm{Tr}H_\gamma(\gamma'-\gamma).
\end{split}
\end{equation}
We thus deduce from above that
$$
\mathrm{Tr}H_\gamma\gamma'\geq \mathrm{Tr}H_\gamma\gamma\ \ \ \mathrm{for\ any}\ \gamma'\in\mathcal{K}_{\lambda},
$$
which implies that $\gamma$ is an optimizer of \eqref{2.8}. Furthermore, if  $\gamma^*$ is a minimizer of the problem \eqref{2.8},  then substituting it into  \eqref{5} yields that
\begin{equation*}
\begin{split}
	\mathcal{E}_\alpha(\gamma)\leq\mathcal{E}_\alpha(\gamma^*)= \mathcal{E}_\alpha(\gamma)-\frac{\alpha^{2p-2}}{p}\int_{\R^3}\Big[\rho_{\gamma^*}^{p}-\rho_\gamma^{p}-p\rho_\gamma^{p-1}
	(\rho_{\gamma^*}-\rho_\gamma)\Big]dx \leq\mathcal{E}_\alpha(\gamma),
\end{split}
\end{equation*}
which gives that $\gamma^*$ is also a minimizer of  $E_\alpha(\lambda)$. This proves the above claim.

We next claim that $H_\gamma$ has at least $N$ non-positive eigenvalues $\mu_1\leq\mu_2\leq\cdots\le\mu_N\leq0$, counted with multiplicity, and the operator
\begin{equation}\label{form}
	\sum_{i=1}^{N-1}|u_i\rangle\langle u_i|+(\lambda-N+1)|u_N\rangle\langle u_N|
\end{equation}
is an optimizer of the problem \eqref{2.8}, where $u_1, \cdots, u_N$ satisfying $(u_i,u_j)=\delta_{ij}$ are the corresponding eigenfunctions of $\mu_1,\cdots, \mu_N$.
To address the claim, one can first verify from \eqref{1} below that there exists a constant $r\geq2$ such that $V(x)-\alpha^{2p-2}\rho_{\gamma}^{p-1}\in L^r(\R^3)+L^\infty_\varepsilon(\R^3)$, where  $L_\varepsilon^\infty(\R^3):=\{\psi\in L^\infty(\R^3): \psi\ \mathrm{approaches\ zero \ at\ infinity}\}$. Following this, we conclude from \cite[Theorem XIII.15]{modern4} that
$\sigma_{ess}(H_\gamma)=\sigma_{ess}(-\Delta)=[0,+\infty)$.   Suppose that $H_\gamma$ has $M$ non-positive eigenvalues $\mu_1\leq\mu_2\leq\cdots\le\mu_M$.
If $0$ is an eigenvalue of $H_\gamma$, then $0$ is infinitely multiple according to \cite[Theorem VII.11]{modern1}, and hence $M=+\infty$. We now assume that $0$ is not an eigenvalue of $H_\gamma$, and let $u_1, \cdots, u_M$  be the corresponding eigenfunctions of $\mu_1, \mu_2, \cdots, \mu_M$, where $(u_i, u_j)=\delta_{ij},\ i,j=1,\cdots,M$. If $M<N$,
by utilizing the min-max theorem (cf. \cite[Theorem 12.1]{analysis}),
one can  check that
\begin{align}\label{ei}
\inf\limits_{\gamma'\in\mathcal{K}_{\lambda}}\mathrm{Tr}H_\gamma(\gamma')
=&\inf\Big\{\sum_{i\geq1}n_i(H_\gamma\varphi_i, \varphi_i): \,  n_i\in[0,1], \  \sum_{i\geq1}n_i=\lambda,\Big.\nonumber\\
&\ \ \ \ \ \ \ \ \Big.\varphi_i\in H^1(\R^3), \ (\varphi_i,\varphi_j)=\delta_{ij} \Big\}=\sum_{i=1}^M\mu_i,
\end{align}
and the infimum cannot be achieved, due to the fact that $0=\min\sigma_{ess}(H_\gamma)$ is not an eigenvalue. This leads to a contradiction, which implies that  $M\geq N$. By \eqref{ei} and the definition of eigenvalues, the operator defined in \eqref{form} is obviously an optimizer of \eqref{2.8}. This proves that the above claim also holds true.

By above two claims, and applying again the definition of eigenvalues or  the \emph{aufbau principle} in quantum chemistry (see \cite{i}), it is now standard to establish \eqref{2.10} and \eqref{6}.
Moreover, note from \cite[Lemma 11.8]{analysis} that  the first eigenfunction $u_1>0$ of $H_\gamma$ is unique, which then indicates that $\mu_1<\mu_2$.
Define $\gamma'=\gamma-t|u_N\rangle\langle u_N|$, where $0<t\leq\lambda-N+1$. It then follows from \eqref{6} and \eqref{5} that
$$
E_\alpha(\lambda-t)\leq \mathcal{E}_\alpha(\gamma')\leq\mathcal{E}_\alpha(\gamma)-t\mu_N=E_\alpha(\lambda)-t\mu_N.
$$
Applying Lemma \ref{lem:2.2}, this gives that $\mu_N\leq t^{-1}E^{\infty}_\alpha(t)$, where $E^\infty_\alpha(t)$  is defined by (\ref{D:1.2}). For any $\hat{\gamma}\in \mathcal{K}_{t}$, we get that
\begin{eqnarray*}
	E^\infty_\alpha(t)\leq\mathcal{E}^\infty_\alpha(\hat{\gamma}_a)=a^2\mathrm{Tr}(-\Delta\hat{\gamma})
	-a^{3(p-1)}\frac{\alpha^{2p-2}}{p}\int_{\R^3}\rho_{\hat{\gamma}}^pdx<0,
\end{eqnarray*}	
if $a>0$ is sufficiently small, where $\hat{\gamma}_a(x,y):=a^3\hat{\gamma}(ax,ay)$ and $\hat{\gamma}(x,y)$ denotes the integral kernel of $\hat{\gamma}$. This further yields that $\mu_N<0$, and Lemma \ref{lem2.3} (1) is thus proved.

2. For any fixed $\alpha>0$, let $\gamma=\sum_{i=1}^{N-1}|u_i\rangle\langle u_i|+(\lambda-N+1)|u_N\rangle\langle u_N|$ be a minimizer of $E_\alpha(\lambda)$  in the form of \eqref{2.10}. We  first claim that
\begin{equation}\label{1}
	u_i\in C(\R^3)\ \ \ \mbox{and}\ \ \  \lim\limits_{|x|\rightarrow\infty}u_i(x)=0.
\end{equation}
Actually, using Kato's inequality {\cite[Theorem X.27]{modern}}, we derive from \eqref{1:1A} and  \eqref{6} that
\begin{equation}\label{a1}
(-\Delta-c_\alpha(x))|u_i|\leq0,\ \  \mbox{where} \ \ c_\alpha(x)=\sum_{k=1}^K|x-y_k|^{-1}+\alpha^{2p-2}\rho_{\gamma}^{p-1}.
\end{equation}
We can further obtain from H\"{o}lder's  inequality that there exists $r\in (3/2,3)$ such that for any $p\in(1,5/3)$,
$$\|c_\alpha\|_ {L^r(B_2(y))}\leq C_1+\alpha C_2\|\rho_\gamma\|_3^{r(p-1)}\ \ \mathrm{holds\  for\  any}\ \  y\in\R^3,$$
where $C_1, C_2>0$ are independent of $\alpha>0$ and $\gamma$.
 Therefore,  applying De Giorgi-Nash-Moser theory (see {\cite[Theorem 4.1]{hq}}) to \eqref{a1}, we deduce that
\begin{equation}\label{a2}
	\|u_i\|_{L^\infty(B_1(y))}\leq C\|u_i\|_{L^2(B_2(y))}\ \ \mathrm{for\ any}\  y\in\R^3,
\end{equation}
where the constant $C>0$ depends  on $\alpha>0$ and $\|\rho_{\gamma}\|_{3}$. This further implies that for fixed $\alpha>0$, we have $u_i\in L^{\infty}(\R^3)$, and hence  $(V(x)-\alpha^{2p-2}\rho_{\gamma}^{p-1})u_i\in L^r_{loc}(\R^3)$ with $r\in(\frac{3}{2},3)$. Consequently, applying the $L^p$ theory \cite{elli}, we derive from \eqref{6} that $u_i(x)\in W^{2,r}_{loc}(\R^3)$.  Combining this with \eqref{a2}, the claim \eqref{1} follows immediately from  Sobolev's embedding theorem.

It follows from \eqref{6} that
\begin{equation}\label{2.1}
\begin{split}
-\Delta\rho_{\gamma}
=&\ 2\sum_{i=1}^{N-1}\left(u_i(-\Delta u_i)-|\nabla u_i|^2\right)+2(\lambda-N+1)\left(u_N(-\Delta u_N)-|\nabla u_N|^2\right)\nonumber\\
\leq&\ 2\sum_{i=1}^{N-1}\left(\mu_iu_i^2+\alpha^{2p-2}
\rho_{\gamma}^{p-1}u_i^2-V(x)u_i^2\right)\nonumber\\
&+2(\lambda-N+1)\left(\mu_Nu_N^2+\alpha^{2p-2}
\rho_{\gamma}^{p-1}u_N^2-V(x)u_N^2\right)\nonumber\\[1.5mm]
\leq&\ 2\big(\mu_N+\alpha^{2p-2}\rho_{\gamma}^{p-1}-V(x)\big)\rho_{\gamma}.
\end{split}
\end{equation}
Since  $\lim\limits_{|x|\to\infty}u_i(x)=0$ and $\lim\limits_{|x|\to\infty}V(x)=0$, there exists a sufficiently large constant $R=R(\alpha)>0$ such that $$\alpha^{2p-2}\rho_{\gamma}^{p-1}(x)-V(x)<-\frac{1}{2}\mu_N\ \  \mbox{for any}\ \ |x|>R,$$
which further implies that
\begin{equation}\label{7}
	\big(-\Delta-\mu_N\big)\rho_{\gamma}(x)\leq0\ \ \mathrm{in}\ \ \R^3\backslash B_R.
\end{equation}
Applying the comparison principle to \eqref{7} then yields that for above sufficiently large $R>0$,
\begin{equation}\label{7A}
	\rho_{\gamma}(x)\leq Ce^{-\sqrt{|\mu_N|}|x|}\ \ \mbox{in}\ \  \R^3\backslash B_R.
\end{equation}
Furthermore, since
\begin{align*}
	\left(-\Delta+V(x)-\alpha^{2p-2}\rho_{\gamma}^{p-1}-\mu_1\right)u_1=0\ \ \mathrm{in}\ \ \R^3,\ \ \ u_1>0,
\end{align*}
and
\begin{align*}
	\left(-\Delta+V(x)-\alpha^{2p-2}\rho_{\gamma}^{p-1}-\mu_i\right)|u_i|\leq0\ \ \mathrm{in}\ \ \R^3,\ \ i=2, \cdots, N,
\end{align*}
other bounds of \eqref{11} and \eqref{2.7} can be  obtained similarly by applying the comparison principle, together with the exponential decay (\ref{7A}). This completes the proof of Lemma 2.3.\qed

\subsection{Proof of Theorem \ref{th1}}

The main purpose of this subsection is to establish Theorem \ref{th1}. One can note from Remark \ref{rem2:2} and Lemma \ref{lem2.3} that for any $\alp>0$ and $N\in\mathbb{N}^+$,
if $E_\alpha(N)$ admits minimizers, then $J_\alpha(N)$ also admits  minimizers and any minimizer  $(u_1, \cdots, u_N)$ of $J_\alpha(N)$ is a ground state of the system \eqref{j1}. In order to establish Theorem \ref{th1}, in this subsection it therefore suffices to
prove the existence of minimizers for $E_\alpha(N)$, instead of $J_\alpha(N)$.
We first have the following strict binding inequality.

\begin{lem}\label{lem2.4}
For any fixed $\alpha>0$, if both $E_\alpha(\lambda_1)$ and $ E_\alpha^\infty(\lambda_2)$ have  minimizers for $ \lambda_1>0$ and $\lambda_2>0$, then we have
\begin{equation*}\label{stric}
	E_\alpha(\lambda_1+\lambda_2)<E_\alpha(\lambda_1)+E^\infty_\alpha(\lambda_2),
\end{equation*}
\end{lem}

Since the proof of Lemma \ref{lem2.4} is similar to that of \cite[Proposition 20]{i},  for the reader's convenience,  we shall
sketch the proof of Lemma \ref{lem2.4} in Appendix A. Applying the above several lemmas, we are now ready to prove Theorem \ref{th1}.

\vspace{.20cm}

\noindent\textbf{Proof of Theorem \ref{th1}.} For any given $\alpha>0$ and $N\in\mathbb{N}^+$, let $\{\gamma_n\}$ be a minimizing sequence of $E_\alpha(N)$.  We can assume from Remark \ref{rem2:2} that there exist $\{u_i^n\}^\infty_{n=1}\subset H^1(\R^3)$ with $(u_i^n, u_j^n)=\delta_{ij}$ such that $\gamma_n=\sum_{i=1}^N|u_i^n\rangle\langle u_i^n|$, where $ i, j=1,\cdots,N$.  Choose $\varepsilon_1>0$ small enough so that
$$
0<\frac{\alpha^{2p-2}}{p}\varepsilon_1<\frac{1}{4}.
$$
Applying Young's inequality, there exists a constant $C_{\varepsilon_1}>0$ such that for $1<p<\frac{5}{3}$,
\begin{align}\label{bound1}
\rho_{n}^p:=\rho_{\gamma_n}^p\leq C_{\varepsilon_1}\rho_n+\varepsilon_1 c_{LT}\rho_n^{\frac{5}{3}},
\end{align}
where $c_{\mathrm{LT}}>0$ is the Lieb-Thirring constant given by \eqref{lt}. By  the inequality (\ref{2:hardy}), we have
\[V(x)=-\sum _{k=1}^K|x-y_k|^{-1}\geq-\varepsilon_2K(-\Delta)-4K\varepsilon_2^{-1}\ \ \mbox{in}\,\ \R^3\,\ \mbox{for any}\,\ \varepsilon_2>0.\]
Choosing $\varepsilon_2>0$ so that $\varepsilon_2K=\frac{1}{2}$, we then have
\begin{align}\label{bound2}
	\mathcal{E}_\alpha(\gamma_n)
	\geq\frac{1}{2}\mathrm{Tr}(-\Delta \gamma_n)-\frac{\alpha^{2p-2}}{p}\int_{\R^3}\rho_{n}^pdx-8K^2N.
\end{align}
By the Lieb-Thirring inequality \eqref{lt}, we therefore get from \eqref{bound1} and \eqref{bound2} that
$$
\mathcal{E}_\alpha(\gamma_n)\geq\frac{1}{4}\mathrm{Tr}(-\Delta \gamma_n)-\frac{\alpha^{2p-2}}{p}C_{\varepsilon_1} N-8K^2N,
$$
which implies that $\{\mathrm{Tr}(-\Delta\gamma_n)\}$ is bounded uniformly for all $n>0$, and hence  $\{u_i^n\}_{n=1}^{\infty}$ is also bounded uniformly in $H^1(\R^3)$ for all $n>0$ and $i=1,\cdots,N$.  Thus, there exist a subsequence, still
denoted by $\{u_i^n\}_{n=1}^\infty$,  of $\{u_i^n\}_{n=1}^\infty$ and $u_i\in H^1(\R^3)$ such that for $i=1,\cdots, N,$
\begin{eqnarray}\label{s1A}
u_i^n\rightharpoonup  u_i\ \ \mathrm{weakly\ in}\  H^1(\R^3) \ \mathrm{as} \ n\to\infty,	
\end{eqnarray}
and
\begin{eqnarray}\label{s1}
|u_i^n|^2\rightarrow  u_i^2 \ \ \mbox{and} \ \ \rho_{\gamma_n}\rightarrow \rho_\gamma\ \   \mathrm{strongly\ in} \ L^r_{loc}(\R^3)\ \mathrm{as} \ n\to\infty,\ \ 1\leq r<3,
\end{eqnarray}
where $\gamma=\sum_{i=1}^N|u_i\rangle\langle u_i|$.
We next proceed the proof by the following three steps:


{\em Step 1.} In this step,  we claim  that $\int_{\R^3}\rho_\gamma dx>0$. By contradiction, suppose $\int_{\R^3}\rho_\gamma dx=0$. It then follows from \eqref{s1} that
\begin{equation*}
	\begin{split}
&\lim\limits_{n\to\infty}\int_{\R^3}V(x)\rho_{\gamma_n}dx
=-\sum_{k=1}^K\lim\limits_{n\to\infty}\int_{\R^3}|x-y_k|^{-1}\rho_{\gamma_n}dx
=0.
\end{split}
\end{equation*}
This gives that
\begin{equation}\label{2.6}
\begin{split}
E_\alpha(N)= \lim\limits_{n\to\infty}\mathcal{E}_\alpha(\gamma_n)=\lim\limits_{n\to\infty}\mathcal{E}^\infty_\alpha(\gamma_n) \geq E_\alpha^\infty(N),
\end{split}
\end{equation}
where $E_\alpha^{\infty}(N)$  is defined by (\ref{D:1.2}).

On the other hand, since $1<p<5/3$, a standard scaling argument gives that $E_\alpha^{\infty}(N)<0$. Let $\{\tilde{\gamma}_n\}=\big\{\sum_{i=1}^N|v_i^n\rangle\langle v_i^n|\big\}$ be a minimizing sequence of  $E_\alpha^{\infty}(N)$, where $(v_i^n, v_j^n)=\delta_{ij}$ for $i, j=1, \cdots, N$. Using the uniform boundedness of $\{v_i^n\}_n$ in $H^1(\R^3)$ and the fact $E_\alpha^{\infty}(N)<0$, we can deduce from the vanishing lemma (cf. \cite[Lemma 1.21]{minmax}) that there exist a constant $R>0$ and a sequence $\{z_n\}\subset\R^3$ such that up to a subsequence if necessary,
\begin{equation}\label{vani}
\lim\limits_{n\to\infty}\int_{B_R(z_n)}\rho_{\tilde{\gamma}_{n}}dx=\lim\limits_{n\to\infty}\sum_{i=1}^N\int_{B_R(z_n)}|v_i^n|^2dx>0,
\end{equation}
and thus there exists a function $\tilde{\rho}\in L^1(\R^3)\backslash\{0\}$ such that
\begin{equation}\label{1:1}
\rho_{\tilde{\gamma}_{n}}(x+z_n)\to\tilde{\rho}\neq0\ \ \text{strongly\ in}\ L_{loc}^1(\R^3) \ \text{as} \ n\to\infty.
\end{equation}
Denote $\tilde{\gamma}^1_n:=\sum_{i=1}^N\big|v_i^n(\cdotp+z_n)\big\rangle\big\langle v_i^n(\cdotp+z_n)\big|$. This then implies from \eqref{1:1} and  Fatou's lemma that
\begin{equation*}\label{j}
\begin{split}
E_\alpha^\infty(N)
=\lim\limits_{n\to\infty}\mathcal{E}^\infty_\alpha(\tilde{\gamma}^1_n)
=&\lim\limits_{n\to\infty}\Big[\mathcal{E}_\alpha(\tilde{\gamma}^1_n)+\sum_{k=1}^K\int_{\R^3}|x-y_k|^{-1}\rho_{\tilde{\gamma}^1_n}dx\Big]\\
\geq& E_\alpha(N)+\sum_{k=1}^K\int_{\R^3}|x-y_k|^{-1}\tilde{\rho}dx>E_\alpha(N),
\end{split}
\end{equation*}
which however  contradicts with \eqref{2.6}. Therefore, the claim $\int_{\R^3}\rho_\gamma dx>0$ holds true.

{\em Step 2.} In this step, we prove  that $\int_{\R^3}\rho_\gamma dx=N$. On the contrary, suppose that $0<\lambda:=\int_{\R^3}\rho_\gamma dx<N$.
Applying an adaptation of the classical dichotomy result,  then there exist a subsequence, still denoted by $\{\rho_n\}$, of $\{\rho_n\}$ and a sequence $\{R_n\}$, where $R_n\rightarrow\infty$ as $n\to\infty$, such that
\begin{equation}\label{17}
0<\lim\limits_{n\to\infty}\int_{|x|\leq R_n}\rho_ndx=\int_{\R^3}\rho_\gamma dx<N,\ \  \lim\limits_{n\to\infty}\int_{R_n\leq|x|\leq 6R_n}\rho_ndx=0.
\end{equation}
Choose a cut-off function $\chi\in C_0^\infty(\R^3)$ satisfying $0\leq\chi\leq1$, where $\chi(x)=1$ for $|x|\leq1$, and $\chi(x)=0$ for $|x|\geq2$. Denote $\chi_{R_n}(x):=\chi(x/R_n)$, $\eta_{R_n}(x):=\sqrt{1-\chi_{R_n}^2(x)}$, and
\begin{equation}\label{4.0}
\begin{split}
 u_i^{1n}:=\chi_{R_n}u_i^n,&\ \ \ \   u_i^{2n}:=\eta_{R_n}u_i^n, \\
\gamma_{1n}:=\sum_{i=1}^{N}|u_i^{1n}\rangle \langle u_i^{1n}|, &\ \ \ \  \gamma_{2n}:=\sum_{i=1}^{N}|u_i^{2n}\rangle \langle u_i^{2n}|.
\end{split}
\end{equation}
For simplicity we denote $\rho_{jn}:=\rho_{\gamma_{jn}}$, $ j=1,2$. We then have
\begin{equation}\label{14}
\begin{split}
 \int_{\R^3}V(x)\rho_{n}dx
 &=\int_{\R^3}V(x)\rho_{1n}dx+\int_{\R^3}V(x)\rho_{2n}dx\\
 &=\int_{\R^3}V(x)\rho_{1n}dx+o(1) \ \ \mbox{as}\ \ n\to\infty.
\end{split}
\end{equation}

Recall from \cite[Theorem 3.2]{ims} that $$-\Delta=\chi_{R_n}(-\Delta)\chi_{R_n}+\eta_{R_n}(-\Delta)\eta_{R_n}-|\nabla\chi_{R_n}|^2-|\nabla\eta_{R_n}|^2.$$
It then yields that
\begin{equation}\label{2.30}
\begin{split}
\mathrm{Tr}(-\Delta\gamma_n)=&\ \mathrm{Tr}(-\Delta\gamma_{1n})+\mathrm{Tr}(-\Delta\gamma_{2n})
-\int_{\R^3}(|\nabla\chi_{R_n}|^2+|\nabla\eta_{R_n}|^2)\rho_ndx\\
\geq&\ \mathrm{Tr}(-\Delta\gamma_{1n})+\mathrm{Tr}(-\Delta\gamma_{2n})
-CR_n^{-2}N,
\end{split}
\end{equation}
where $C>0$ is independent of $n>0$.
As for the nonlinear term, we rewrite
$$
\rho_n=\chi_{R_n}^2\rho_{n}+\eta_{R_n}^2\chi_{3R_n}^2\rho_n+\eta_{3R_n}^2\rho_n.
$$
It follows from \eqref{17} that $\eta_{R_n}^2\chi_{3R_n}^2\rho_n\rightarrow0$ strongly in $L^1(\R^3)$ as $n\to\infty$. By the uniform boundedness of $\{\rho_n\}$ in $L^{\frac{5}{3}}(\R^3)$, we then conclude that $\eta_{R_n}^2\chi_{3R_n}^2\rho_n\rightarrow0$ strongly in $L^p(\R^3)$ as $n\to\infty$, and hence
\begin{equation}\label{16}
\begin{split}
\int_{\R^3}\rho_{n}^pdx=&\int_{\R^3}\left(\chi_{R_n}^2\rho_{n}
+\eta_{3R_n}^2\rho_{n}\right)^pdx+o(1) \\
=& \int_{\R^3}\Big[\left(\chi_{R_n}^2\rho_{n}\right)^p
+\left(\eta_{3R_n}^2\rho_{n}\right)^p\Big]dx+o(1)\\
=& \int_{\R^3}\Big[\left(\chi_{R_n}^2\rho_{n}\right)^p
+\left(\eta_{R_n}^2\chi_{3R_n}^2\rho_n+\eta_{3R_n}^2\rho_{n}\right)^p\Big]dx+o(1)\\
=&\int_{\R^3}\left(\rho_{1n}^p+\rho_{2n}^p\right)dx+o(1)\ \ \mathrm{as}\ \ n\to\infty.
\end{split}
\end{equation}

Since $\lim\limits_{n\rightarrow\infty}\int_{\R^3}\rho_{1n}dx=\lambda$ and $\lim\limits_{n\rightarrow\infty}\int_{\R^3}\rho_{2n}dx=N-\lambda$, applying Lemma \ref{lem:2.2} (2), we  conclude from \eqref{14}--\eqref{16} that
\begin{equation}\label{19}
\begin{split}
&\ E_\alpha(\lambda)+E^\infty_\alpha(N-\lambda)\geq E_\alpha(N)=\lim\limits_{n\to\infty}\mathcal{E}_\alpha(\gamma_n)\\[1mm]
 \geq&\lim\limits_{n\to\infty}\mathcal{E}_\alpha(\gamma_{1n})+\lim\limits_{n\to\infty}\mathcal{E}^\infty_\alpha(\gamma_{2n})
\geq E_\alpha(\lambda)+E^\infty_\alpha(N-\lambda),
\end{split}	
\end{equation}
where  the continuities of $E_\alpha(\cdot)$ and $E^\infty_\alpha(\cdot)$ are employed.   This thus yields that
\begin{align}\label{2.42a}
\lim\limits_{n\rightarrow\infty}\mathcal{E}_\alpha(\gamma_{1n})=E_\alpha(\lambda)\ \ \text{and}\ \ \lim\limits_{n\rightarrow\infty}\mathcal{E}^\infty_\alpha(\gamma_{2n})=E^\infty_\alpha(N-\lambda).
\end{align}
Moreover, it follows from \eqref{s1A} and  \eqref{17} that the sequences  $\{u_i^{1n}\}_{n=1}^\infty$ and $\{\gamma_{1n}\}$ defined in \eqref{4.0}  satisfy
$$
u_i^{1n}\rightharpoonup u_i\ \ \mathrm{weakly\ in}\ \ H^1(\R^3),
\ \  \rho_{1n}=\sum_{i=1}^N|u_i^{1n}|^2\rightarrow\rho_\gamma=\sum_{i=1}^Nu_i^2\ \ \mathrm{strongly\ in}\ \ L^1(\R^3)
$$
as $ n\to\infty$. Using the interpolation inequality and the boundedness of $\{\rho_{1n}\}$ in $L^{3}(\R^3)$, we further have
\begin{equation*}
	\rho_{1n}\rightarrow\rho_\gamma\ \ \text{strongly\ in}\ \ L^r(\R^3)\ \ \text{as}\ \ n\to\infty,\ \ r\in[1, 3).
\end{equation*}
This implies from \eqref{2.42a} that
\begin{equation}\label{13a}
\gamma\ \text{is\ a\ minimizer\ of}\  E_\alpha(\lambda).
\end{equation}

On the other hand, it yields from \eqref{2.42a} that once  $\int_{\R^3}\rho_\gamma dx:=\lambda\in(0, N)$, then $\{\gamma_{2n}\}$ is a minimizing sequence of $E_\alpha^\infty(N-\lambda)$. We next consider the following two cases:

Case 1:  $\{\rho_{\gamma_{2n}}\}$ is relatively compact, up to a subsequence and translations if necessary. In this case, one can get that $E_\alpha^\infty(N-\lambda)$ possesses at least one minimizer. Using this and \eqref{13a}, we then  deduce from Lemma \ref{lem2.4} that
\begin{equation*}
E_\alpha(N)<E_\alpha(\lambda)+E^\infty_\alpha(N-\lambda),
\end{equation*}
which however contradicts with \eqref{19}.  Therefore, this completes the proof of Step 2.

Case 2:  $\{\rho_{\gamma_{2n}}\}$ is not relatively compact, up to a subsequence and translations. In this case, the same argument of proving \eqref{vani} then gives that  the sequence $\{\rho_{\gamma_{2n}}\}$ cannot vanish. Accordingly, similar to \eqref{4.0}, up to a subsequence and translations if necessary, we can decompose the sequence $\{\gamma_{2n}\}$ into two sequences  $\big\{\gamma^{(1)}_{2n}\big\}$ and $\big\{\gamma^{(2)}_{2n}\big\}$.  The same arguments of proving \eqref{19} and \eqref{13a} further give that there exists $\lambda_2\in(0, N-\lambda)$ such that
\begin{equation}\label{2.45}
\begin{split}
E^\infty_\alpha(N-\lambda)
=\lim\limits_{n\to\infty}\mathcal{E}_\alpha^\infty(\gamma_{2n})
&=\lim\limits_{n\to\infty}\mathcal{E}_\alpha^\infty(\gamma_{2n}^{(1)})+\lim\limits_{n\to\infty}\mathcal{E}_\alpha^\infty(\gamma_{2n}^{(2)})\\
&=E^\infty_\alpha(\lambda_2)+E^\infty_\alpha(N-\lambda-\lambda_2),
\end{split}
\end{equation}
and
\begin{align}\label{2.44}
E^\infty_\alpha(\lambda_2)\ \text{\ admits\ at\ least\ one\ minimizer}.
\end{align}
Combining \eqref{19} and \eqref{2.45}, we then obtain that
\begin{align}\label{2.46}
E_\alpha(N)=E_\alpha(\lambda)+E^\infty_\alpha(\lambda_2)+E^\infty_\alpha(N-\lambda-\lambda_2).
\end{align}
However, by \eqref{13a} and \eqref{2.44},  we deduce from Lemmas \ref{lem2.4} and \ref{lem:2.2} (2) that
$$
E_\alpha(\lambda)+E^\infty_\alpha(\lambda_2)+E^\infty_\alpha(N-\lambda-\lambda_2)>E_\alpha(\lambda+\lambda_2)+E^\infty_\alpha(N-\lambda-\lambda_2)\geq E_\alpha(N),
$$
which unfortunately contradicts with \eqref{2.46}.  Therefore, this also  completes the proof of Step 2.

{\em Step 3.}   The previous two steps now yield that $\rho_n\rightarrow \rho_\gamma$ in $L^1(\R^3)$ as $n\to\infty$, and hence $u_i^n\rightarrow u_i$ in $L^2(\R^3)$ as $n\to\infty$, where $ i=1,\cdots, N$.  Using the interpolation inequality and the uniform boundedness of $\{\rho_n\}$ in $L^{3}(\R^3)$, we further have
\begin{equation*}\label{13}
	\rho_n\rightarrow\rho_\gamma\ \ \text{strongly\ in}\ \ L^r(\R^3)\ \ \text{as}\ \ n\to\infty,\ \ r\in[1, 3).
\end{equation*}
Consequently, by weak lower semicontinuity, we deduce that
$$
 E_\alpha(N)=\liminf\limits_{n\to\infty}\mathcal{E}_\alpha(\gamma_n)\geq\mathrm{Tr}(-\Delta\gamma)+\int_{\R^3}V(x)\rho_\gamma dx-\frac{\alpha^{2p-2}}{p}\int_{\R^3}\rho_\gamma^pdx\geq E_\alpha(N),$$
which implies that $\gamma$ is a minimizer of $E_\alpha(N)$, and we are therefore done.\qed

	
\section{Limiting Behavior of Minimizers as $\alpha\to\infty$}

This section is devoted to analyzing the limiting behavior of  minimizers for $J_\alpha(N)$ as $\alpha\rightarrow\infty$, where  $N\in\mathbb{N}^+$ is fixed, and the potential $V(x)<0$ is as in \eqref{1:1A}.
Following Theorem \ref{th1} and \cite[Theorem 4]{i}, there exists a constant $p_c\in(1,\frac{5}{3}]$ such that for any $p\in(1, p_c)$, both $J_1^\infty(N)$  and  $J_{\alpha}(N)$ admit minimizers for all $\alpha>0$,
where  $J_{\alpha}(N)$ is given by \eqref{1.0}, and  $J_1^\infty(N)$  is defined as
\begin{gather}\label{3:jinfty}
	J_1^\infty(N):=\inf\Big\{\mathcal{E}_1^\infty(u_1,\cdots,u_N):\, u_1,\cdots,u_N\in H^1(\R^{3}),\ (u_i,u_j)_{L^2}=\delta_{ij}\Big\}.
\end{gather}
Here the energy functional $\mathcal{E}_1^\infty(u_1,\cdots,u_N)$ satisfies
\begin{gather*}
	\mathcal{E}_1^\infty(u_1,\cdots,u_N)=\sum_{i=1}^N\int_{\R^3}|\nabla u_i|^2dx-\frac{1}{p}\int_{\R^3}\Big(\sum_{i=1}^N| u_i|^2\Big)^pdx.
\end{gather*}
Throughout this section we always assume $p\in(1,p_c)$, where $p_c\in(1,\frac{5}{3}]$ is given by Theorem \ref{th1}.

In this section, we always denote    $(\hat{w}_1,\cdots,\hat{w}_N)$ and $(u_1^n,\cdots,u_N^n)$  a  minimizer  of $J_{1}^\infty(N)$ and  $J_{\alpha_n}(N)$,  respectively, where  $\alpha_n\to\infty$ as $n\to\infty$. Set
\begin{equation}\label{3.02A}
\hat{\gamma}:=\sum_{i=1}^N|\hat{w}_i\rangle\langle \hat{w}_i|,\ \ \ \gamma_n:=\sum_{i=1}^{N}|u_i^n\rangle\langle u_i^n|.
\end{equation}	
We also define for $i=1,\cdots,N$,
\begin{equation}\label{3.02}
\epsilon_n:=\alpha_n^{\frac{-2(p-1)}{2-3(p-1)}}>0, \ \   w_i^n(x):=\epsilon_n^{\frac{3}{2}}u_i^n(\epsilon_n x),\ \   \tilde{\gamma}_n:=\sum_{i=1}^{N}|w_i^n\rangle\langle w_i^n|,
\end{equation}	
so that $\epsilon_n\to 0$ as  $n\to\infty$, and
\begin{eqnarray}\label{3.1}
\epsilon_n^2\mathrm{Tr}(-\Delta\gamma_n)=\mathrm{Tr}(-\Delta\tilde{\gamma}_n),\ \ \epsilon_n^2\alpha_n^{2p-2}\int_{\R^3}\rho_{\gamma_n}^pdx=\int_{\R^3}\rho_{\tilde{\gamma}_n}^pdx,
\end{eqnarray}
where $\rho_{\gamma_n}\!=\sum_{i=1}^N|u_i^n|^2$ and $\rho_{\tilde{\gamma}_n}\!=\sum_{i=1}^N|w_i^n|^2$  are defined by \eqref{2.2D}--\eqref{2.2E}. We start with the following energy estimates as $n\to\infty$.

\begin{lem}\label{lem:3.1}
Suppose $\gamma_n$ is defined by (\ref{3.02A}), and let $\epsilon_n> 0$ be as in (\ref{3.02}). Then there exist some constants $M_1>M_2>0$, $M'_1>M'_2>0$, $M''_1>M''_2>0$ and $M'''_1>M'''_2>0$, which are independent of $n>0$, such that for sufficiently large $n>0$,
\begin{gather}\label{3.5}
M_2\leq \epsilon_n^2\mathrm{Tr}(-\Delta\gamma_n)\leq M_1,\ \ \ M'_2\leq \epsilon_n^2\alpha_n^{2p-2}\|\rho_{\gamma_n}\|_p^p\leq M'_1, \\[1mm]
M_2''\leq -\epsilon_n\int_{\R^3}V(x)\rho_{\gamma_n}dx\leq M''_1,\ \ M'''_2\epsilon_n\leq J_{1}^\infty(N)-\epsilon_n^{2}J_{\alpha_n}(N)\leq M'''_1\epsilon_n,
\label{3.7}
\end{gather}
where $p\in(1,p_c)$ and  $p_c\in(1,\frac{5}{3}]$ is given by Theorem \ref{th1}.
\end{lem}

\noindent \textbf{Proof.} Define
$$\hat{w}^n_i(x):=\epsilon_n^{-\frac{3}{2}}\hat{w}_i(\epsilon^{-1}_n x), \ \ i=1, 2, \cdots, N,$$
where $\epsilon_n>0$ is as in \eqref{3.02}, and $(\hat{w}_1,\cdots,\hat{w}_N)$ is a minimizer of  $J_{1}^\infty(N)$ defined in \eqref{3:jinfty}.
By scaling, it is easy to check that  $(\hat{w}_1^n,\cdots,\hat{w}_N^n)$ is a minimizer of $J_{\alpha_n}^\infty(N)$ and $J_{\alpha_n}^\infty(N)=J_{1}^\infty(N)\epsilon_n^{-2}$, where $J_{\alpha_n}^\infty(N)$ is given by \eqref{jinfty}. We thus obtain from \eqref{bound2} and  \eqref{3.1} that for all $n\ge 1$,
\begin{align}\label{3.0}
	0>J_{1}^\infty(N)=&\ \epsilon_n^2J_{\alpha_n}^\infty(N)\geq\epsilon_n^2J_{\alpha_n}(N)
	=\epsilon_n^2\mathcal{E}_{\alpha_n}(\gamma_n)\nonumber\\[1mm]
	\geq&\ \epsilon_n^2\Big(\frac{1}{2}\mathrm{Tr}(-\Delta\gamma_n)
	-\frac{\alpha_n^{2p-2}}{p}\int_{\R^3}\rho_{\gamma_n}^pdx-8NK^2\Big)\nonumber\\
	=&\ \frac{1}{2}\mathrm{Tr}(-\Delta\tilde{\gamma}_n)
	-\frac{1}{p}\int_{\R^3}\rho_{\tilde{\gamma}_n}^pdx-8NK^2\epsilon_n^2\\
	\geq&\ \frac{1}{2}\mathrm{Tr}(-\Delta\tilde{\gamma}_n)	 -\frac{1}{p}\Big(K^{-1}(p,N)\mathrm{Tr}(-\Delta\tilde{\gamma}_n)\Big)^{\frac{3(p-1)}{2}}-8NK^2\epsilon_n^2,\nonumber
\end{align}
where $\tilde{\gamma}_n$ is given by (\ref{3.02}), and  the last inequality follows from the Gagliardo-Nirenberg-Sobolev inequality  \eqref{gn}.
Since $0<\frac{3(p-1)}{2}<1$, we derive from \eqref{3.0} that  $\mathrm{Tr}(-\Delta\tilde{\gamma}_n)$ is bounded uniformly in $n>0$. Applying  Hoffmann-Ostenhof  inequality \eqref{ho}, we deduce that $\|\rho_{\tilde{\gamma}_n}\|_p$ is also bounded uniformly in $n>0$, which thus gives the upper bounds of \eqref{3.5}. The lower bounds of \eqref{3.5} follow directly from \eqref{gn} and  \eqref{3.0}. Actually,  if $\mathrm{Tr}(-\Delta\tilde{\gamma}_n)=o(1)$ as $n\to\infty$, then we obtain from  \eqref{gn} that $\|\rho_{\tilde{\gamma}_n}\|_p=o(1)$ as $n\to\infty$. Combining this with \eqref{3.0}, one gets that $0>J_{1}^\infty(N)\geq0$, a contradiction. This implies that the sequence $\{\mathrm{Tr}(-\Delta\tilde{\gamma}_n)\}$ has a positive lower bound. Similarly,  using \eqref{3.0} again, we conclude that if $\|\rho_{\tilde{\gamma}_n}\|_p=o(1)$ as $n\to\infty$, then
$$
0>J_{1}^\infty(N)\geq \frac{1}{2}\mathrm{Tr}(-\Delta\tilde{\gamma}_n)+o(1)>0\ \
\mathrm{as}\ \ n\to\infty.
$$
This shows that the sequence $\{\|\rho_{\tilde{\gamma}_n}\|_p\}$  has also a positive lower bound, and \eqref{3.5} hence holds true.

We next prove \eqref{3.7}.  Since we can get from \eqref{2:hardy} that
$$
-V(x)\leq\epsilon_n K(-\Delta)+4\epsilon_n^{-1}K\ \ \mathrm{  in}\  \R^3,
$$
where $K\in\mathbb{N}^+$ is given in \eqref{1:1A}, it implies that
\begin{equation}\label{3.01}
\begin{split}
-\epsilon_n^2\int_{\R^3}V(x)\rho_{\gamma_n}dx
&\leq  \epsilon_n^2\big[ \epsilon_n K\mathrm{Tr}(-\Delta \gamma_n)+4\epsilon_n^{-1}KN\big]\\
&\leq  \epsilon_n \big(K M_1+4NK\big),
\end{split}
\end{equation}
where $M_1>0$ is given by \eqref{3.5}.
This proves  the upper bound of  $-\epsilon_n\int_{\R^3}V(x)\rho_{\gamma_n}dx$ as $n\to\infty$.
As for its lower bound, 
by contradiction,  suppose that $-\epsilon_n\int_{\R^3}V(x)\rho_{\gamma_n}dx= o(1)$ as $n\rightarrow\infty$. It then yields from \eqref{3.1} that
\begin{equation}\label{3.8}
\begin{split}	\epsilon_n^2J_{\alpha_n}(N)
	&=\mathrm{Tr}(-\Delta\tilde{\gamma}_n)
	-\frac{1}{p}\int_{\R^3}\rho_{\tilde{\gamma}_n}^pdx
	+\epsilon_n^2\int_{\R^3}V(x)\rho_{\gamma_n}dx\\
	&\geq J_{1}^\infty(N)-o(\epsilon_n)\ \ \mathrm{as}\ \ n\rightarrow\infty.
\end{split}
\end{equation}
On the other hand, letting $\hat{\gamma}_n:=\sum_{i=1}^N|\hat{w}^n_i\rangle\langle \hat{w}^n_i|$, it then gives  that
\begin{equation}\label{3.8.0}
\begin{split}
\epsilon_n^2J_{\alpha_n}(N)&\leq  \epsilon_n^2\mathcal{E}_{\alpha_n}\big(\hat{\gamma}_n(\cdot-y_1)\big)
\\
&=\epsilon_n^2\Big[\mathrm{Tr}(-\Delta\hat{\gamma}_n)-\frac{\alpha_n^{2p-2}}{p}\int_{\R^3}
\rho_{\hat{\gamma}_n}^pdx 	+\int_{\R^3}V(x)\rho_{\hat{\gamma}_n}(x-y_1)dx\Big] \\
&=  \epsilon_n^2\Big[J_{\alpha_n}^\infty(N)-\epsilon_n^{-1}\int_{\R^3}\sum_{k=1}^K
\left|x+\epsilon_n^{-1}(y_1-y_k)\right|^{-1}\rho_{\hat{\gamma}}(x)dx\Big] \\
&\leq J_{1}^\infty(N)-C\epsilon_n\ \ \mathrm{as}\ \ n\rightarrow\infty,
\end{split}
\end{equation}
where $y_k$ is as in \eqref{1:1A}, and $C>0$ is independent of $n$. This however contradicts with \eqref{3.8}. Together with \eqref{3.01}, this  thus yields the bounds of $-\epsilon_n\int_{\R^3}V(x)\rho_{\gamma_n}dx$ as $n\to \infty$.

By the upper bound of $-\epsilon_n\int_{\R^3}V(x)\rho_{\gamma_n}dx$ as $n\to \infty$, we finally derive from \eqref{3.1} that
\begin{align*}
	\epsilon_n^2J_{\alpha_n}(N)
	&=\mathrm{Tr}(-\Delta\tilde{\gamma}_n)
	-\frac{1}{p}\int_{\R^3}\rho_{\tilde{\gamma}_n}^pdx
	+\epsilon_n^2\int_{\R^3}V(x)\rho_{\gamma_n}dx\\
	&\geq J_{1}^\infty(N)+\epsilon_n^2\int_{\R^3}V(x)\rho_{\gamma_n}dx\\
	&\geq J_{1}^\infty(N)-M''_1\epsilon_n\ \ \mathrm{as}\ \ n\rightarrow\infty.
\end{align*}
Together with \eqref{3.8.0}, this further gives the bounds of $\epsilon_n^2J_{\alpha_n}(N)$ as $n\to \infty$, and Lemma \ref{lem:3.1} is therefore proved.\qed

We next establish the limiting behavior and the uniformly  decaying  result of minimizers  for $J_{\alpha_n}(N)$ with $\alpha_n\to\infty$ as $n\to\infty$.  Let $(u_1^n,\cdots,u_N^n)$ be a minimizer of $J_{\alpha_n}(N)$ satisfying
\begin{align}\label{3nls}
H_V^nu_i^n=\mu_i^n u_i^n\ \ \mathrm{in}\ \R^3, \ \ i=1,\cdots,N,
\end{align}
where
\begin{align}\label{hnv}
	H_V^n:=-\Delta+V(x)-\alpha_n^{2p-2}\Big(\sum_{j=1}^N|u_j^n|^2\Big)^{p-1},
\end{align}
and $\mu_1^n<\mu_2^n\leq\cdots\leq \mu_N^n<0$ are the $N$ first  eigenvalues  of  the operator $H_V^n$ in $\R^3$, counted with multiplicity. We point out that for any $n>0$, the function $\rho_{\gamma_n}=\sum_{i=1}^N|u_i^n|^2$  has at least one global maximum point  in view of \eqref{1}. Moreover,  recall from \cite{i} that if the minimizer  $(\hat{w}_1,\cdots,\hat{w}_N)$  of $J^\infty_1(N)$ solves the following fermionic system
\begin{align}\label{3nls1}
-\Delta \hat{w}_i-\Big(\sum_{j=1}^N\hat{w}_j^2\Big)^{p-1}\hat{w}_i=\hat{\mu}_i \hat{w}_i\ \ \mathrm{in}\ \R^3,\ \ i=1,\cdots,N,
\end{align}
then $\hat{\mu}_1<\hat{\mu}_2\leq\cdots\leq \hat{\mu}_N<0$ are the $N$ first eigenvalues  of $-\Delta-\big(\sum_{j=1}^N\hat{w}_j^2\big)^{p-1}$  in $\R^3$, counted with multiplicity.

Following above notations, we now establish the following convergence.

\begin{lem}\label{lem:3.2} Let  $(u_1^n,\cdots,u_N^n)$ be a minimizer of $J_{\alpha_n}(N)$  with $\alpha_n\to\infty$ as $n\to\infty$, and suppose  $\mu_i^n$ is the $i$th  eigenvalue (counted with multiplicity) of the operator $H_V^n$ defined by \eqref{hnv} in $\R^3$, which corresponds to the eigenfunction $u_i^n$ for $i=1,\cdots,N$. Then the following statements hold:
\begin{enumerate}
\item  There exists a subsequence, still denoted by  $\{(u_1^n,\cdots,u_N^n)\}$, of  $\{(u_1^n,\cdots,u_N^n)\}$  such that for $i=1,\cdots,N$,
\begin{equation}\label{3.9}
\hat{w}_i^n:=\epsilon_n^{\frac{3}{2}}u_i^n(\epsilon_n\cdot+z_n)\to \hat{w}_i\ \ \text{strongly\ in}\ \ H^1(\R^3)\ \ \mathrm{as}\ \ n\to\infty,
\end{equation}
where $\epsilon_n=\alpha_n^{\frac{-2(p-1)}{2-3(p-1)}}>0$ is as in \eqref{3.02}, $z_n$ is a global maximum point of the function $\rho_{\gamma_n}=\sum_{i=1}^N|u_i^n|^2$, and  $(\hat{w}_1,\cdots, \hat{w}_N)$ is a minimizer of $J_1^\infty(N)$.
Moreover, there exists some $y_{k}\in\{y_1,\cdots,y_K\}$ given in  \eqref{1:1A} such that the  global maximum point  $z_n$ of $\rho_{\gamma_n}$  satisfies \begin{equation}\label{3.9a}
|z_n-y_{k}|\leq C\epsilon_n\ \ as\ \  n\to\infty,
\end{equation}
where $C>0$ is independent of $n>0$.
\item The $i$th eigenvalue $\mu_i^n$ of the operator $H_V^n$  in $\R^3$ satisfies
\begin{equation}\label{def1}
\lim\limits_{n\rightarrow\infty}\epsilon_n^2\mu_i^n=\hat{\mu}_i<0,\ \ \ i=1,\cdots, N,
\end{equation}
where $\hat{\mu}_i$ is the $i$th  (counted with multiplicity) eigenvalue of 	 $-\Delta-\big(\sum_{j=1}^N\hat{w}_j^2\big)^{p-1}$ in $\R^3$, which corresponds to the eigenfunction $\hat{w}_i$ for $i=1,\cdots,N$.
\end{enumerate}
\end{lem}

\noindent \textbf{Proof.} 1. By the definition of $\tilde{\gamma}_n=\sum_{i=1}^{N}|w_i^n\rangle\langle w_i^n|
$ in \eqref{3.02}, we derive from Lemma \ref{lem:3.1} that
$$
J_{1}^\infty(N)-O(\epsilon_n)= \epsilon_n^2J_{\alpha_n}(N)
=\mathrm{Tr}(-\Delta\tilde{\gamma}_n)
-\frac{1}{p}\int_{\R^3}\rho_{\tilde{\gamma}_n}^pdx-O(\epsilon_n)\ \ \mathrm{as}\  n\to\infty,
$$
which implies that $\{(w_1^n,\cdots,w_N^n)\}_n$ is a minimizing sequence of $J_{1}^\infty(N)$. Therefore, following \cite[Theorem 27]{geomrtric} and \cite[Theorem 3]{i}, up to a subsequence,  there exist a sequence $\{\bar{z}_{n}\}\subset \R^3$ and   a minimizer $(\bar{w}_1,\cdots,\bar{w}_N)$
of $J_{1}^\infty(N)$ such that
\begin{equation}\label{3.13a}
\bar{w}_i^{n}:=w_i^{n}(\cdot+\bar{z}_{n})=\epsilon_n^{\frac{3}{2}}u_i^n\big(\epsilon_n(\cdot+\bar{z}_{n})\big)\rightarrow \bar{w}_i\ \mathrm{\  strongly\ in}\ H^1(\R^3)\ \mathrm{as}\  n\rightarrow\infty,
\end{equation}
where $\epsilon_n=\alpha_n^{\frac{-2(p-1)}{2-3(p-1)}}>0$ is given by \eqref{3.02}.

Denote $\bar{\gamma}_n:=\sum_{i=1}^{N}|\bar{w}_i^n\rangle\langle \bar{w}_i^n|$ and $\bar{\gamma}:=\sum_{i=1}^N|\bar{w}_i\rangle\langle \bar{w}_i|$. We now claim that the sequence $\{\epsilon_n^{-1}|z_n-\epsilon_n\bar{z}_n|\}$ is bounded, where $z_n$ is a global maximum point of  
$\rho_{\gamma_n}=\sum_{i=1}^N|u_i^n|^2$.  Actually,  we deduce from \eqref{3nls} that $\bar{w}_i^n$  is a solution of
\begin{eqnarray}\label{3.2}
	-\Delta\bar{w}_i^n+\epsilon_n^2V\big(\epsilon_n(\cdot+\bar{z}_n)\big)\bar{w}_i^n
	-\rho_{\bar{\gamma}_n}^{p-1}\bar{w}_i^n=\epsilon_n^2\mu_i^n \bar{w}_i^n\ \ \mathrm{in}\ \R^3,\ \ i=1,\cdots,N.
\end{eqnarray}
Hence, using  the boundedness of the sequences $\{\bar{w}_i^n\}_{n=1}^\infty$ in $H^1(\R^3)$, $i=1,\cdots,N$, the same argument of \eqref{a2} gives that for sufficiently large $n>0$,
\begin{equation}\label{3.10}
	\|\rho_{\bar{\gamma}_n}\|_{L^\infty(B_1(y))}\leq C\|\rho_{\bar{\gamma}_n}\|_{L^1(B_2(y))}\ \ \mathrm{for\ any}\  y\in\R^3,
\end{equation}
where $C>0$ is  independent of $n>0$ and $y$.
This further indicates that
\begin{equation}\label{3.10b}
	\lim\limits_{|x|\to\infty}\rho_{\bar{\gamma}_n}(x)=0 \ \ \mathrm{uniformly\ for\ sufficiently\ large}\ n>0.
\end{equation}
Since $\frac{z_n-\epsilon_n\bar{z}_n}{\epsilon_n}$ is a global
maximum point of $\rho_{\bar{\gamma}_n}=\sum_{i=1}^{N}|\bar{w}_i^n|^2$, we deduce from \eqref{3.10b} that if $|\frac{z_n-\epsilon_n\bar{z}_n}{\epsilon_n}|\rightarrow\infty$, then $\rho_{\bar{\gamma}_n}\rightarrow0$ a.e. in $\R^3$, which is impossible in view of the fact that  $\rho_{\bar{\gamma}_n}\rightarrow\rho_{\bar{\gamma}}:=\sum_{i=1}^N\bar{w}_i^2>0$ a.e. in $\R^3$ as $n\to\infty$. The above claim therefore holds true.

Hence, 
up to a subsequence if necessary, we can assume that $\epsilon_n^{-1}(z_n-\epsilon_n\bar{z}_n)\to z_*$ as $n\to\infty$ for some $z_*\in\R^3$. It then follows from \eqref{3.13a} that
\begin{equation}\label{3.19a}
\begin{split}
\hat{w}_i^n(x):=&\ \epsilon_n^{\frac{3}{2}}u_i^n(\epsilon_nx+z_n)=\bar{w}_i^n\big(x+\epsilon_n^{-1}(z_n-\epsilon_n\bar{z}_n)\big)\\[1mm]
\rightarrow &\ \bar{w}_i\big(x+z_*\big):=\hat{w}_i(x)\ \ \  \mathrm{strongly\ in}\  H^1(\R^3)  \ \ \mathrm{as}\ n\to\infty.
\end{split}
\end{equation}
By the translation invariance of the minimization problem $J_1^\infty(N)$,  we deduce from \eqref{3.19a}  that $(\hat{w}_1, \cdots, \hat{w}_N)$ is also a minimizer of  $J_1^\infty(N)$. This implies immediately that \eqref{3.9} holds true.

We next claim that there exists  some $k\in\{ 1, 2,\cdots, K\}$  such that the sequence $\{\epsilon_n^{-1}|\epsilon_n\bar{z}_n-y_{k}|\}_{n=1}^\infty$ is bounded, where $y_k\in\R^3$ is given in \eqref{1:1A}.
Actually, by contradiction, without loss of generality, we assume that $\lim\limits_{n\rightarrow\infty}\epsilon_n^{-1}|\epsilon_n\bar{z}_n-y_k|\rightarrow\infty$ for any $y_k$, $k=1,\cdots, K$. We then  conclude from \eqref{3.13a} that
\begin{align*}
	0\leq&-\lim\limits_{n\to\infty}\epsilon_n\int_{\R^3}V(\epsilon_n x+\epsilon_n \bar{z}_n)\rho_{\bar{\gamma}_n}(x)dx\\
	=&\lim\limits_{n\to\infty}\sum_{k=1}^{K}\int_{\R^3}\big|x-\epsilon_n^{-1}(y_k -\epsilon_n\bar{z}_n)\big|^{-1}\rho_{\bar{\gamma}_n}(x)dx\\
	 =&\lim\limits_{n\to\infty}\sum_{k=1}^{K}\int_{\R^3}\big|x-\epsilon_n^{-1}(y_k-\epsilon_n\bar{z}_n)\big|^{-1}\rho_{\bar{\gamma}}(x)dx\\
	 =&\lim\limits_{R\to\infty}\lim\limits_{n\to\infty}\sum_{k=1}^{K}\int_{B_R(\epsilon_n^{-1}(y_k-\epsilon_n\bar{z}_n))}\big|x-\epsilon_n^{-1}(y_k-\epsilon_n\bar{z}_n)\big|^{-1}\rho_{\bar{\gamma}}(x)dx\\
	 &+\lim\limits_{R\to\infty}\lim\limits_{n\to\infty}\sum_{k=1}^{K}\int_{B_R^c(\epsilon_n^{-1}(y_k-\epsilon_n\bar{z}_n))}\big|x-\epsilon_n^{-1}(y_k-\epsilon_n\bar{z}_n)\big|^{-1}\rho_{\bar{\gamma}}(x)dx\\
	 \leq&\lim\limits_{R\to\infty}\lim\limits_{n\to\infty}\sum_{k=1}^{K}\int_{B_R(0)}|x|^{-1}\rho_{\bar{\gamma}}\big(x+\epsilon_n^{-1}(y_k-\epsilon_n\bar{z}_n)\big)dx\\
	 &+\lim\limits_{R\to\infty}\lim\limits_{n\to\infty}\sum_{k=1}^{K}\frac{1}{R}\int_{\R^3}\rho_{\bar{\gamma}}dx\\
	 \leq&\lim\limits_{R\to\infty}\lim\limits_{n\to\infty}\sum_{k=1}^{K}\big\||x|^{-1}\big\|_{L^{\frac{7}{3}}(B_R(0))}\Big(\int_{B_R(0)}\rho_{\bar{\gamma}}^{\frac{7}{4}}\big(x+\epsilon_n^{-1}(y_k-\epsilon_n\bar{z}_n)\big)dx\Big)^{\frac{4}{7}}\\
	=&0,
\end{align*}
where $\rho_{\bar{\gamma}}=\sum_{i=1}^N\bar{w}_i^2$, and the last identity follows from the fact that $\rho_{\bar{\gamma}}\in L^r(\R^3)$ holds for $r\in[1,3]$. We thus derive that
\begin{align*}\label{3.4}
	\epsilon_n^2J_{\alpha_n}(N)=&\epsilon_n^2\Big(\mathrm{Tr}(-\Delta\gamma_n)
	-\frac{\alpha^{2p-2}}{p}\int_{\R^3}\rho_{\gamma_n}^pdx+\int_{\R^3}V(x)\rho_{\gamma_n}dx\Big)\\
	=&\ \mathrm{Tr}(-\Delta\bar{\gamma}_n)
	-\frac{1}{p}\int_{\R^3}\rho_{\bar{\gamma}_n}^pdx+\epsilon_n^2\int_{\R^3}V(\epsilon_n x+\epsilon_n \bar{z}_n)\rho_{\bar{\gamma}_n}dx\nonumber\\
	\geq&\ J_{1}^\infty(N)+o(\epsilon_n)\ \ \mathrm{as}\ \ n\to\infty,
\end{align*}
which however contradicts with \eqref{3.7}. Therefore, the claim holds true.

It is obvious that
\begin{equation}\label{3.22b}
\epsilon_n^{-1}|z_n-y_{k}|\leq\epsilon_n^{-1}|z_n-\epsilon_n\bar{z}_n|+\epsilon_n^{-1}|\epsilon_n\bar{z}_n-y_{k}|.
\end{equation}
By the above boundedness of the sequences $\{\epsilon_n^{-1}|z_n-\epsilon_n\bar{z}_n|\}$ and  $\{\epsilon_n^{-1}|\epsilon_n\bar{z}_n-y_{k}|\}_{n=1}^\infty$, we then conclude from \eqref{3.22b} that $|z_n-y_{k}|\leq C\epsilon_n$ as  $n\to\infty$, where $C>0$ is independent of $n>0$. This implies that \eqref{3.9a} holds true.

2. One can get from \eqref{3.2} and \eqref{3.19a} that $\hat{w}_i^n$ satisfies
\begin{eqnarray}\label{3.16}
	-\Delta\hat{w}_i^n+\epsilon_n^2V(\epsilon_n\cdot+z_n)\hat{w}_i^n
	-\rho_{\hat{\gamma}_n}^{p-1}\hat{w}_i^n=\epsilon_n^2\mu_i^n \hat{w}_i^n\ \ \mathrm{in}\ \R^3,\ \ i=1,\cdots, N.
\end{eqnarray}
Together with \eqref{3.9}, we then conclude that
\begin{equation*}
\begin{split}
	\sum_{i=1}^{N}\epsilon_n^2\mu_i^n
	&=\mathrm{Tr}(-\Delta\hat{\gamma}_n)
	+\epsilon_n^2\int_{\R^3}V(\epsilon_n x+z_n)\rho_{\hat{\gamma}_n}dx
	-\int_{\R^3}\rho^p_{\hat{\gamma}_n}dx\\
	&=\mathrm{Tr}(-\Delta\hat{\gamma})-\int_{\R^3}\rho^p_{\hat{\gamma}}dx+o(1)
	\ \ \mathrm{as}\ \ n\to\infty,
\end{split}
\end{equation*}
where $\hat{\gamma}_n=\sum_{i=1}^N|\hat{w}_i^n\rangle\langle \hat{w}_i^n|$ and $\hat{\gamma}=\sum_{i=1}^N|\hat{w}_i\rangle\langle \hat{w}_i|$. This further indicates that for $ i=1,\cdots,N$, the sequence $\{\epsilon_n^2\mu_i^n\}_{n=1}^\infty$ is also bounded in view of  the fact that $\mu_1^n<\mu_2^n\leq\cdots\leq \mu_N^n<0$. Therefore, up to a subsequence if necessary, we can assume that
$0\geq\lambda_i:=\lim\limits_{n\rightarrow\infty}\epsilon_n^2\mu_i^n$, where $i=1,\cdots,N$. Moreover, we note from \eqref{3.9} and \eqref{3.16} that $\hat{w}_i$ satisfies
$$
-\Delta \hat{w}_i-\rho_{\hat{\gamma}}^{p-1}\hat{w}_i=\lambda_i \hat{w}_i\ \ \mathrm{in}\ \ \R^3,\ \ i=1,\cdots,N.
$$
This shows from \eqref{3nls1} that $\lambda_i=\hat{\mu}_i$ for $i=1,\cdots,N$, where $\hat{\mu}_i$ is the $i$th  (counted with multiplicity) eigenvalue of 	$-\Delta-\big(\sum_{j=1}^N\hat{w}_j^2\big)^{p-1}$ in $\R^3$, which corresponds to the eigenfunction $\hat{w}_i$ for $i=1,\cdots,N$, and  we are therefore done. 
\qed

We next establish the exponential decay of the sequence $\{\hat{w}_i^n(x)\}_{n=1}^\infty$ given by Lemma \ref{lem:3.2}, where $i=1,\cdots, N$. Note from \eqref{3.16} that $\hat{w}_i^n$ satisfies the following NLS system
\begin{eqnarray}\label{3.24}
	\big(-\Delta+\epsilon_n^2|\mu_i^n|\big) \hat{w}_i^n=\epsilon_n\sum_{k=1}^K\big|x-\epsilon_n^{-1}(y_k-z_n)\big|^{-1}\hat{w}_i^n
	+\rho_{\hat{\gamma}_n}^{p-1}\hat{w}_i^n\ \ \mathrm{in}\,\ \R^3,
\end{eqnarray}
where $i=1,\cdots, N$, where $z_n\in\R^3$ is given by Lemma \ref{lem:3.2} (1), and  $-\infty<\epsilon_n^2\mu_i^n<0$ holds for all $n>0$ in view of \eqref{def1}. However,  one cannot apply the standard comparison principle to yielding the exponential decay of $\hat{w}_i^n(x)$ as $|x|\to\infty$, due to the singularity of the Coulomb potential in \eqref{3.24}. We shall employ  the Green's  function to establish the following exponential decay of $\hat{w}_i^n(x)$ as $|x|\to\infty$.

\begin{lem}\label{lem:3.3}
Suppose that  $\hat{w}_i^n$  and $\hat{\mu}_i<0$ are given by Lemma \ref{lem:3.2}, where $i=1,\cdots, N$.   
Then for any $\theta_i\in\big(0,\sqrt{|\hat{\mu}_i|}\big)$, there is a constant $C(\theta_i)>0$, independent of $n>0$, such that for sufficiently large $n>0$,
	\begin{equation*}\label{3.04}
		|\hat{w}_i^n(x)|\leq C(\theta_i)e^{-\theta_i|x|}\ \ in\ \ \R^3,\ \ i=1, \ \cdots,\ N.
	\end{equation*}
\end{lem}

\noindent{\bf Proof.}  Fix $\theta_i^*\in(0, \sqrt{|\hat{\mu}_i|})$ and denote $\rho_n:=\rho_{\hat{\gamma}_n}=\sum_{i=1}^N|\hat{w}_i^n|^2$, where $\hat{\gamma}_n=\sum_{i=1}^N|\hat{w}_i^n\rangle\langle \hat{w}_i^n|$. 
It yields from \eqref{3.24}   that
$$
\hat{w}_i^n(x)=\int_{\R^3}G_i^n(x-y)\Big(\epsilon_n\sum_{k=1}^{K}|y-\epsilon_n^{-1}( y_k-z_n)|^{-1}+\rho_n^{p-1}(y)\Big)\hat{w}_i^n(y)dy,
$$
where $G_i^n(x)$ is the Green's function of $-\Delta-\epsilon_n^2\mu_i^n$ in $\R^3$, $i=1,\cdots,N$. Note from \cite[Theorem 6.23]{analysis} 
that
$$
G_i^n(x)=\frac{1}{4\pi|x|}e^{-\sqrt{|\epsilon_n^2\mu_i^n|}|x|}\ \ \mathrm{in}\ \ \R^3.
$$
Therefore,  it follows from \eqref{def1} that  for sufficiently large $n>0$,
\begin{equation}\label{3.00}
\begin{split}
	|\hat{w}_i^n(x)|\leq\ &C\int_{\R^3}|x-y|^{-1}e^{{-\theta_i^*}|x-y|}|\hat{w}_i^n(y)|\\
	&\cdot\Big(\epsilon_n\sum_{k=1}^{K}|y-\epsilon_n^{-1} (y_k-z_n)|^{-1}+\rho_n^{p-1}(y)\Big)dy\ \ \mathrm{in}\,\ \R^3,
\end{split}
\end{equation}
where $C>0$ is independent of $i>0$ and $n>0$.
Inspired by the Slaggie-Wichmann method in \cite{expo}, we define for $i=1,\cdots,N$,
\begin{equation}\label{3.20a}
\begin{split}
	m_i^n(x):=&\sup\limits_{y\in\R^3}\Big\{|\hat{w}_i^n(y)|e^{-\theta_i|x-y|}\Big\},
\end{split}
\end{equation}
and
\begin{equation}\label{3.20b}
	\begin{split}
	h^n_{i,\theta}(x):=&C\int_{\R^3}|x-y|^{-1}
	e^{-(\theta_i^*-\theta_i)|x-y|}\rho_n^{p-1}(y)dy\\
	&+C\epsilon_n\sum_{k=1}^{K}\int_{\R^3}|x-y|^{-1}
	e^{-(\theta_i^*-\theta_i)|x-y|}|y-\epsilon_n^{-1} (y_k-z_n)|^{-1}dy\\
	=&:I^n_{\rho}(x)+I^n_V(x),
\end{split}
\end{equation}
where  $0<\theta_i<\theta_i^*<\sqrt{|\hat{\mu}_i|}$. It then  follows from \eqref{3.00} that for sufficiently large $n>0$,
\begin{equation}\label{3.18}
	|\hat{w}_i^n(x)|\leq m_i^n(x)h^n_{i,\theta}(x)\ \ \ \mathrm{in}\,\ \R^3,\ \ i=1,\cdots,N.
\end{equation}

We first claim that there exist a constant $C(\theta_i)>0$ and a sufficiently large $R >0$, independent of $n>0$, such that  for sufficiently large $n>0$,
\begin{equation}\label{claim1}
	\ \ h^n_{i,\theta}(x)<C(\theta_i)\ \ \mathrm{in}\ \, \R^3,
\end{equation}
and
\begin{equation}\label{claim2}
	h^n_{i,\theta}(x)<\frac{1}{2}\ \ \mathrm{in}\ \ \R^3\backslash B_R,
\end{equation}
where $	h^n_{i,\theta}$ is defined by \eqref{3.20b}.
Indeed, as for the term $I^n_{\rho}(x)$, using the boundedness of $\{\sqrt{\rho_{n}}\}$ in $H^1(\R^3)$, we deduce from \eqref{3.10}--\eqref{3.19a} that
\begin{equation}\label{3.30}
	 \sup\limits_{n>0}\|\rho_n\|_{\infty}=\sup\limits_{n>0}\big\|\Sigma_{i=1}^N|\hat{w}_i^n|^2\big\|_{\infty}<+\infty,
\end{equation}
and
\begin{equation}\label{3.22a}
		\lim\limits_{|x|\to\infty}\rho_{n}(x)=0\
\ \mathrm{uniformly\ for \ sufficiently\ large}\ n>0.
\end{equation}
Thus, one can easily check that   for sufficiently large $n>0$,
\begin{equation}\label{3.11}
	I^n_{\rho}(x):=C\int_{\R^3}|x-y|^{-1}
	e^{-(\theta_i^*-\theta_i)|x-y|}\rho_n^{p-1}(y)dy\leq C(\theta_i)\ \ \mathrm{in}\,\  \R^3,
\end{equation}	
where $C(\theta_i)>0$ depends only on $\theta_i>0$. Furthermore, 
one gets  from \eqref{3.22a} that for any sufficiently  small  $\varepsilon>0$, there exists $R'_\varepsilon >2$ such that  for sufficiently large $n>0$,
$$
\rho_n^{p-1}(x)<\varepsilon\ \ \mathrm{ in}\,\  \R^3\backslash B_{R'_\varepsilon}(0).
$$
Thus for above sufficiently small $\varepsilon>0$, there exists $R_\varepsilon:=\max\{2R'_\varepsilon, \ \varepsilon^{-2}\}$ such that for any sufficiently large $n>0$,
\begin{equation}\label{3.19}
\begin{split}
I^n_{\rho}(x)
\leq&\ C|x|^{-\frac{1}{2}}
\int_{|x-y|\geq|x|^{\frac{1}{2}}}e^{-(\theta_i^*-\theta_i)|x-y|}\rho_n^{p-1}(y)dy\\[1mm]
&+C\int_{|x-y|<|x|^{\frac{1}{2}}}|x-y|^{-1}e^{-(\theta_i^*-\theta_i)|x-y|}
\rho_n^{p-1}(y)dy\\[1mm]
\leq&\ C\|\rho_n\|^{p-1}_{\infty}|x|^{-\frac{1}{2}}
\int_{\R^3}e^{-(\theta_i^*-\theta_i)|y|}dy
+\varepsilon C\int_{|z|<|x|^{\frac{1}{2}}}
|z|^{-1}e^{-(\theta_i^*-\theta_i)|z|}dz\\[1mm]
<&\ \varepsilon C(\theta_i)\ \ \mathrm{in}\ \R^3\backslash B_{R_\varepsilon}(0).
\end{split}
\end{equation}
As for the term $I^n_V(x)$, we obtain from H\"{o}lder's inequality that for sufficiently large $n>0$,
\begin{eqnarray}\label{3.20}
\begin{split}
	I^n_V(x):=&\ C\epsilon_n\sum_{k=1}^K\int_{\R^3}|x-y|^{-1}|y-\epsilon_n^{-1} (y_k-z_n)|^{-1}e^{-(\theta_i^*-\theta_i)|x-y|}dy\\
	\leq&\ C\epsilon_n\left\||x-y|^{-1}e^{-\frac{1}{2}(\theta_i^*-\theta_i)|x-y|}\right\|_{2}\\
	 &\cdot\sum_{k=1}^K\left\||y-\epsilon_n^{-1}(y_k-z_n)|^{-1}e^{-\frac{1}{2}(\theta_i^*-\theta_i)|x-y|}\right\|_{2}	 \ \ \ \mathrm{in}\  \R^3.		 
\end{split}
\end{eqnarray}
We therefore derive from \eqref{3.20b} and \eqref{3.11}--\eqref{3.20} that  the above  claim  is true.

By  the boundedness of $h^n_{i,\theta}$ in \eqref{claim1}, we now deduce from \eqref{3.20a} and \eqref{3.18} that for sufficiently large $n>0$,
\begin{eqnarray}\label{3.21}
	|\hat{w}_i^n(x)|\leq C(\theta_i)m_i^n(x)
	=C(\theta_i)\sup\limits_{y\in\R^3}\left\{|\hat{w}_i^n(y)|e^{-\theta_i|x-y|}\right\}
	\ \ \mathrm{in}\ \ \R^3,\ \ i=1,\cdots,N.
\end{eqnarray}
Moreover, we drive from \eqref{3.18} and \eqref{claim2} that for sufficiently large $n>0$,
$$
|\hat{w}_i^n(y)|<\frac{1}{2}m_i^n(y)\ \ \ \mathrm{in}\  B_R^c, \ \ i=1,\cdots,N,
$$
where $R>0$ is as in  \eqref{claim2}.
This further indicates that for sufficiently large $n>0$,
\begin{align*}
	\sup\limits_{y\in B_R^c}\Big\{|\hat{w}_i^n(y)|e^{-\theta_i|x-y|}\Big\}
	&\leq\frac{1}{2}\sup\limits_{y\in B_R^c}\Big\{m_i^n(y)e^{-\theta_i|x-y|}\Big\}\\
	&\leq\ \frac{1}{2}\sup\limits_{y\in \R^3}\Big\{m_i^n(y)e^{-\theta_i|x-y|}\Big\}\\
	&=\frac{1}{2}\sup\limits_{y\in \R^3}\Big\{\sup\limits_{z\in \R^3}\big\{
	|\hat{w}_i^n(z)|e^{-\theta_i|y-z|}\big\}e^{-\theta_i|x-y|}\Big\}\\
	&=\frac{1}{2}\sup\limits_{z\in \R^3}\Big\{|\hat{w}_i^n(z)|\sup\limits_{y\in \R^3}\big\{
	e^{-\theta_i|y-z|}e^{-\theta_i|x-y|}\big\}\Big\}\\
	&=\frac{1}{2}\sup\limits_{z\in \R^3}\Big\{|\hat{w}_i^n(z)|
	e^{-\theta_i|x-z|}\Big\}\\
	&=\ \frac{1}{2}m_i^n(x)<m_i^n(x)\ \ \ \mathrm{in} \ \R^3,\ \ i=1,\cdots,N.
\end{align*}
By the definition of $m_i^n$, we hence derive  from above that for sufficiently large $n>0$,
\begin{eqnarray*}
	\begin{split}
		m_i^n(x)=&\max\Big\{\sup\limits_{y\in B_R}\big\{|\hat{w}_i^n(y)|e^{-\theta_i|x-y|}\big\},\ \sup\limits_{y\in B_R^c}\big\{|\hat{w}_i^n(y)|e^{-\theta_i|x-y|}\big\}\Big\}\\
		\leq&\sup\limits_{y\in B_R}\Big\{|\hat{w}_i^n(y)|e^{-\theta_i|x-y|}\Big\}\ \ \ \mathrm{in} \ \R^3,\ \ i=1,\cdots,N.
	\end{split}
\end{eqnarray*}
Together with \eqref{3.21}, we then conclude from \eqref{3.30} that
\begin{equation*}\label{3.22}
\begin{split}
	|\hat{w}_i^n(x)|
	&\leq C(\theta_i)m_i^n(x)\leq C(\theta_i)\sup\limits_{y\in B_R}\Big\{|\hat{w}_i^n(y)|e^{-\theta_i|x-y|}\Big\}\\
	&\leq C(\theta_i)e^{\theta_iR}e^{-\theta_i|x|}\sup\limits_{n>0}\|\hat{w}_i^n\|_\infty <C'(\theta_i)e^{-\theta_i|x|}
	\ \ \ \mathrm{in} \ \R^3, \ \ i=1,\cdots,N,
\end{split}
\end{equation*}
where $R>0$ is as in  \eqref{claim2}. This therefore completes the proof of  Lemma \ref{lem:3.3}. \qed

\vspace{.20cm}

\noindent\textbf{Proof  of Theorem \ref{th2}.} Let  $(u_1^n,\cdots,u_N^n)$ be a minimizer of $J_{\alpha_n}(N)$  with $\alpha_n\to\infty$ as $n\to\infty$, and suppose $\hat{w}_i^n(x)=\epsilon_n^{\frac{3}{2}}u_i^n(\epsilon_nx+z_n) $ is defined as in Lemma \ref{lem:3.2}. It then follows from Lemmas \ref{lem:3.2} and  \ref{lem:3.3} that 
 in order to establish Theorem \ref{th2}, we  just need to prove that, up to a subsequence if necessary,
\begin{equation}\label{3.37}
	\hat{w}_i^{n}\to \hat{w}_i\ \ \text{strongly\ in}\ \ L^\infty(\R^3)\ \ \mathrm{as}\ \ n\rightarrow\infty, \ \ i=1,\cdots,N,
\end{equation}
where  $\hat{w}_i$ is defined by Lemma \ref{lem:3.2}.

Recall from \eqref{3.16} that $\hat{w}_i^{n}$ satisfies
\begin{eqnarray*}
	-\Delta\hat{w}_i^n=-\epsilon_n^2V(\epsilon_n\cdot+z_n)\hat{w}_i^n
	+\rho_{\hat{\gamma}_n}^{p-1}\hat{w}_i^n+\epsilon_n^2\mu_i^n \hat{w}_i^n:=f_i^n(x)\ \ \mathrm{in}\ \ \R^3,\ \ i=1,\cdots, N.
\end{eqnarray*}
Since it yields from \eqref{3.9} and  \eqref{3.30} that the sequence $\{\hat{w}_i^{n}\}_{n=1}^\infty$ is bounded uniformly  in $H^1(\R^3)\cap L^\infty(\R^3)$, we obtain that $\{f_i^n(x)\}_{n=1}^\infty$ is  bounded uniformly in $L^{r}_{loc}(\R^3)$, where $r\in(3/2, 3)$ and $i=1,\cdots,N$. Applying the $L^p$ theory to the above system, we thus get that for any fixed $ R>0$,
$$
\|\hat{w}_i^n\|_{W^{2, r}(B_R)}\leq C\Big(\|\hat{w}_i^n\|_{L^{r}(B_{R+1})}
+ \|f_i^n\|_{L^{r}(B_{R+1})}   \Big),\ \  \ i=1,\cdots,N,
$$
where  
$C>0$ is independent of $n>0$. This shows that $\{\hat{w}_i^n\}_{n=1}^\infty$ is bounded uniformly  in $W^{2, r}(B_R)$ for $r\in(3/2, 3)$, $i=1,\cdots,N$. Since the embedding $W^{2,r}(B_R)\hookrightarrow C(B_R)$ is compact (c.f.\cite[Theorem 7.26]{elli}) for $r\in(3/2, 3)$, we deduce from the convergence of \eqref{3.9} that there exists a subsequence, still denoted by $\{\hat{w}_i^{n}\}_{n=1}^\infty$, of $\{\hat{w}_i^{n}\}_{n=1}^\infty$ such that for any  fixed $R>0$,
\begin{equation}\label{conv1}
	\hat{w}_i^{n}\rightarrow\hat{w}_i\ \ \mbox{strongly in}\ \ L^\infty(B_R)\ \ \mathrm{as}\ n\to\infty,\ \ i=1,\cdots,N.
\end{equation}

On the other hand,  note from Lemma \ref{lem:3.2} that $\hat{w}_i\in H^1(\R^3)$ satisfies
\begin{align}\label{3.38a}
	-\Delta \hat{w}_i-\Big(\sum_{j=1}^N\hat{w}_j^2\Big)^{p-1}\hat{w}_i=\hat{\mu}_i \hat{w}_i\ \ \mathrm{in}\ \R^3,\ \ i=1,\cdots, N,
\end{align}
where $\hat{\mu}_i<0$ holds for all  $i=1,\cdots,N$.  Applying the standard elliptic regularity theory \cite{i} to \eqref{3.38a}, it then yields that $\hat{w}_i\in C(\R^3)$ and $\lim\limits_{|x|\to\infty}\hat{w}_i(x)=0$. By the comparison principle, we thus obtain from \eqref{3.38a} that
\begin{align}\label{ex}
|\hat{w}_i(x)|\leq Ce^{-\sqrt{|\hat{\mu}_i|}|x|} \ \ \mathrm{ in}\ \R^3, \ \ i=1,\cdots, N,
\end{align}
where $C>0$. By the exponential decay of Lemma \ref{lem:3.3}, we get from \eqref{ex} that for any $\varepsilon>0$, there exists a sufficiently large constant $R:=R(\varepsilon) > 0$, independent of $n>0$, such that for sufficiently large $n>0$,
\begin{eqnarray*}\label{3.32}
|\hat{w}_i(x)|,\ \ |\hat{w}_i^{n}(x)|<\frac{\varepsilon}{4}\ \ \ \mathrm{in}\ \ \R^3\backslash B_R,\ \ i=1,\cdots,N,
\end{eqnarray*}
and thus,
\begin{eqnarray}\label{3.33}
	\sup\limits_{|x|\geq R}|\hat{w}^n_i(x)-\hat{w}_i(x)|\leq	\sup\limits_{|x|\geq R}\big(|\hat{w}^n_i(x)|+|\hat{w}_i(x)|\big)<\frac{\varepsilon}{2},\ \ i=1,\cdots,N.
\end{eqnarray}


Combining \eqref{conv1} with \eqref{3.33}, we finally conclude that \eqref{3.37} holds true, and Theorem \ref{th1} is therefore proved. \qed
	
\appendix
\section{Appendix}
\renewcommand{\theequation}{A.\arabic{equation}}
\setcounter{equation}{0}	
	
In this Appendix, we first illustrate briefly how to derive the relation (\ref{1.0}), and we then address the proof of Lemma \ref{lem2.4}.

\vspace{.2cm}
\noindent\textbf{Proof  of (\ref{1.0}).}
By  the definition of $J_\alpha(N)$ in (\ref{1.0}), one can get from \eqref{1.0a} that
\begin{equation}\label{1.0b}
\begin{split}
	J_\alpha(N)
	=&\inf\Big\{\mathcal{E}(\Psi): \Psi \ \mathrm{is\ a\ Slater \ determinant}, \ \, \|\Psi\|_{2}=1, \, \ \Psi\in H^1(\R^{3N},\R)\Big\} \\
	\geq&E_f(N).
\end{split}
\end{equation}
On the other hand,  let $\gamma_\Psi$ be the one-particle density matrix associated with $\Psi$, i.e.,
$$
\gamma_\Psi(x,y):=N\int_{\R^{3(N-1)}}\Psi(x,x_2,\cdots,x_N)
\overline{\Psi}(y,x_2,\cdots,x_N)dx_2\cdots dx_N.
$$
One can then check that $\gamma_\Psi\!\in\! \mathcal{B}\big(L^2(\R^3, \C)\big)$, $0\leq\gamma_\Psi=\gamma_\Psi^*\leq 1$, $\mathrm{Tr}\gamma_\Psi=N$ and
\begin{equation*}\label{1.7D}
\mathcal{E}_\alpha(\gamma_\Psi):=\mathrm{Tr}\big(-\Delta+V(x)\big) \gamma_\Psi-\frac{\alpha^{2p-2}}{p}\int_{\R^3}\rho_{\gamma_\Psi}^pdx=\mathcal{E}(\Psi),
\end{equation*}
where $\rho_{\gamma_\Psi}(x)=\gamma_\Psi(x, x)$ and $\mathcal{B}\big(L^2(\R^3, \C)\big)$ denotes the set of bounded linear operators on $L^2(\R^3, \C)$. This  further implies from \eqref{1.0a} that
\begin{equation}\label{1.9}
E_f(N)\geq \inf\limits_{\gamma\in\mathcal{K}'_{N}}\mathcal{E}_\alpha(\gamma),
\end{equation}
where
\begin{equation*}\label{1.11}
		\mathcal{K}'_{N}=:\big\{\gamma\in\mathcal{B}\big(L^2(\R^3, \C)\big):\ 0\leq\gamma=\gamma^*\leq 1,\ \mathrm{Tr}\gamma=N,\ \mathrm{Tr}(-\Delta \gamma)<\infty\big\}.
\end{equation*}
Moreover, since
$\mathcal{E}_\alpha(\gamma)
=\mathcal{E}_\alpha\Big(\frac{\gamma+\bar{\gamma}}{2}\Big)$ and $\frac{\gamma+\bar{\gamma}}{2}$ is a bounded linear operator on $ L^2(\R^3, \R)$ for any $\gamma\in \mathcal{K}'_{N}$, where $\bar{\gamma}$ denotes the complex conjugate of $\gamma$, we deduce from \eqref{1.0b} and \eqref{1.9} that
\begin{equation}\label{1.13}
J_\alpha(N)\geq E_f(N)\geq  \inf\limits_{\gamma\in\mathcal{K}_{N}}\mathcal{E}_\alpha(\gamma),
\end{equation}
where the space $\mathcal{K}_{N}$ is defined as
\begin{equation*}
\mathcal{K}_{N}:=\big\{\gamma\in \mathcal{B}\big(L^2(\R^3, \R)\big):\ 0\leq\gamma=\gamma^*\leq 1,\ \mathrm{Tr}\gamma=N,\ \mathrm{Tr}(-\Delta \gamma)<\infty\big\}.
\end{equation*}

Furthermore, the similar argument of \cite[Lemma 11]{i} yields that
\begin{equation}\label{1.13A}
\inf\limits_{\gamma\in\mathcal{K}_{N}}\mathcal{E}_\alpha(\gamma)=\inf\Big\{  \mathcal{E}_\alpha(\gamma):\gamma=\sum_{i=1}^{N}|u_i\rangle\langle u_i|, u_i\in H^1(\R^3, \R), (u_i,u_j)_{L^2}=\delta_{ij}\Big\},
\end{equation}
where $\gamma=\sum_{i=1}^{N}|u_i\rangle\langle u_i|$ is a bounded linear operator on $ L^2(\R^3, \R)$ and satisfies
$$
\gamma\varphi(x)=\sum_{i=1}^Nu_i(x)(\varphi, u_i) \ \ \text{for any}\ \ \varphi\in L^2(\R^3,\R),
$$
see those around Lemma \ref{lem:2.1} for more details.
Applying  the definition of $J_\alpha(N)$ in \eqref{1.0},
we thus get from (\ref{1.13A}) that
\begin{equation}\label{1.12B}
\inf\limits_{\gamma\in\mathcal{K}_{N}}\mathcal{E}_\alpha(\gamma)=J_\alpha(N).
\end{equation}
We therefore conclude from \eqref{1.13} and (\ref{1.12B}) that $J_\alpha(N)=E_f(N)$, $i.e.,$ (\ref{1.0})  holds true.\qed

\vspace{.20cm}

\noindent \textbf{Proof of Lemma \ref{lem2.4}.} Let
$$\gamma_1:=\sum_{i=1}^{N}|u_i\rangle\langle u_i|+(\lambda_1-N)|u_N\rangle\langle u_N|,$$
and
$$ \gamma_2=:\sum_{j=1}^{M}|v_j\rangle\langle v_j|+(\lambda_2-M)|v_M\rangle\langle v_M|$$
be a minimizer of $E_\alpha(\lambda_1)$ and $ E_\alpha^\infty(\lambda_2)$, respectively, where $N$ and $ M$ are the smallest integers such that $\lambda_1\leq N,  \lambda_2\leq M$.
Define $v_j^R:=v_j(\cdot-Re_1),\ j=1,\cdots, M$,  where $R>0$ and $e_1=(1,0,0)$. Motivated by \cite{i}, we now consider the Gram matrix $G_R$ of the family $u_1, \cdots, u_N, v_1^R, \cdots, v_M^R$, i.e.,
$$
G_R:=\left(
\begin{split}
	\mathbb{I}_N\ \ \ \ \ \ &A^R\\
	(A^R)^*\ \ \ \ \ &\mathbb{I}_M
\end{split}
\right),\ \ A^R=(A^R_{ij})_{N\times M}, \ \ A^R_{ij}=(u_i, v_j^R),
$$
where $\mathbb{I}_{N}$ denotes the $N$-order identity matrix.
By the definition of $G_R$, we deduce that for sufficiently large $R>0$, the matrix $G_R$ is positive definite, and hence
$$
I_{N+M}=G_R^{-\frac{1}{2}}\left(
\begin{split}
	\mathbb{I}_N\ \ \ \ \ \ &A^R\\
	(A^R)^*\ \ \ \ \ &\mathbb{I}_M
\end{split}
\right)G_R^{-\frac{1}{2}}
$$
holds for sufficiently large $R>0$. This shows that for sufficiently large $R>0$, the components of the vector
\begin{equation}\label{2.5}
	(\tilde{u}_1^R, \cdots, \tilde{u}_N^R, \tilde{v}_1^R, \cdots, \tilde{v}_M^R):=(u_1, \cdots, u_N, v_1^R, \cdots, v_M^R)G_R^{-\frac{1}{2}}
\end{equation}
are orthonormal in $L^2(\R^3)$.

Define
\begin{equation*}\label{8}
	\begin{split}
		\gamma_R:=
		\sum_{i=1}^{N}|\tilde{u}_i^R\rangle\langle \tilde{u}_i^R|+(\lambda_1-N)|\tilde{u}_N^R\rangle\langle \tilde{u}_N^R|+\sum_{j=1}^{M}|\tilde{v}_j^R\rangle\langle \tilde{v}_j^R|+(\lambda_2-M)|\tilde{v}_M^R\rangle\langle \tilde{v}_M^R|.
	\end{split}
\end{equation*}
We then derive that $\gamma_R\in \mathcal{K}_{\lambda_1+\lambda_2}$ holds for sufficiently large $R>0$. Since $(1+t)^{-\frac{1}{2}}=1-\frac{1}{2}t+O(t^2)$ as $t\to0$, we have
$$
G_R^{-\frac{1}{2}}=
\left(\begin{split}
	\mathbb{I}_N\ \ \ \ \ \ &0\\
	0\ \ \ \ \ \ \ &\mathbb{I}_M
\end{split}
\right)
-\frac{1}{2}\left(
\begin{split}
	0\ \ \ \ \ \ &A^R\\
	(A^R)^*\ \ \ \ &0
\end{split}
\right)
+O(a_R^2)\left(
\begin{split}
	E_N\ \ \ \ \ \ &0\\
	0\ \ \ \ \ \ \ &E_M
\end{split}
\right)\ \ \mathrm{as}\ \ R\to\infty,
$$
where $a_R=\max\limits_{i,j}\big|(u_i,v_j^R)\big|$, and $E_N$ denotes the  $N$-order matrix with all elements being $1$. As a consequence,
we derive from \eqref{2.5} that
\begin{equation*}
	\begin{split}
		&\ (\tilde{u}_1^R, \cdots, \tilde{u}_N^R, \tilde{v}_1^R, \cdots, \tilde{v}_M^R)
		=\ (u_1, \cdots, u_N, v_1^R, \cdots, v_M^R)\\[1mm]
		-&\frac{1}{2}\Big(\sum_{j=1}^MA_{1j}^Rv_j^R,\cdots,
		\sum_{j=1}^MA_{Nj}^Rv_j^R,\sum_{i=1}^NA_{i1}^Ru_i,\cdots,
		\sum_{i=1}^NA_{iM}^Ru_i\Big)+O(a_R^2)\ \ \mathrm{as}\ \ R\to\infty.
	\end{split}
\end{equation*}
This further implies that
\begin{align}\label{2.9}
		\gamma_R=&\gamma_1+\gamma'_2-\sum_{i=1}^{N}\sum_{j=1}^{M}A^R_{ij}\left(|u_i\rangle\langle v_j^R|+|v_j^R\rangle\langle u _i|\right)\nonumber\\
		&-\frac{1}{2}(\lambda_1-N)\sum_{j=1}^{M}A^R_{Nj}\left(|u_N\rangle\langle v_j^R|+|v_j^R\rangle\langle u _N|\right)\\
		&-\frac{1}{2}(\lambda_2-M)\sum_{i=1}^{N}A^R_{iM}\left(|u_i\rangle\langle v_M^R|+|v_M^R\rangle\langle u _i|\right)+O(a_R^2)\ \ \mathrm{as}\ \ R\to\infty,\nonumber
\end{align}
where $\gamma'_2:=\sum_{j=1}^{M}|v_j^R\rangle\langle v_j^R|+(\lambda_2-M)|v_M^R\rangle\langle v_M^R|$.

We now deduce from \eqref{6} and \eqref{2.9} that
\begin{equation*}
	\begin{split}
		\mathrm{Tr}(-\Delta+V)\gamma_R
		=&\ \mathrm{Tr}(-\Delta+V)(\gamma_1 +\gamma'_2)
		-2\sum_{i=1}^{N}\sum_{j=1}^{M}A^R_{ij}\Big(\big(\alpha^{2p-2}\rho_{\gamma_1}^{p-1}+\mu_i\big)u_i,v_j^R\Big)\\[0mm]
		&+(\lambda_1-N)\sum_{j=1}^{M}A_{Nj}^R\Big(\big(\alpha^{2p-2}
		\rho_{\gamma_1}^{p-1}+\mu_N\big)u_N,v_j^R\Big)\\
		&+(\lambda_2-M)\sum_{i=1}^{N}A_{iM}^R\Big(\big(\alpha^{2p-2}
		\rho_{\gamma_1}^{p-1}+\mu_i\big)u_i,v_M^R\Big)+O(a_R^2)\\
		=&\ \mathrm{Tr}(-\Delta+V)(\gamma_1 +\gamma'_2)+O(a_R^2)\ \ \ \mathrm{as}\ \ R\to\infty,
	\end{split}
\end{equation*}
and
$$
\int_{\R^3}\rho_{\gamma_R}^pdx=\int_{\R^3}\left(\rho_{\gamma_1}+\rho_{\gamma'_2}\right)^pdx+O(a_R^2)\ \ \ \mathrm{as}\ \ R\to\infty,
$$
where  $\mu_i$ denotes the $i$th eigenvalue of the operator $-\Delta+V(x)-\alpha^{2p-2}\rho_{\gamma_1}^{p-1}$. Since $\rho_{\gamma_2}\geq v_1^2>0$, we have
\begin{equation}\label{2.42}
	\begin{split}
		\mathrm{Tr}\big(V(x)\gamma'_2\big)=&\int_{\R^3}-\sum_{k=1}^{K}|x-y_k|^{-1}\rho_{\gamma_2}(x-Re_1)dx\\
		<&-R^{-1}\int_{|x|\leq1}|e_1+ R^{-1}(x-y_1)|^{-1}\rho_{\gamma_2}(x)dx\\
		\leq&-CR^{-1}\ \ \mathrm{as}\ \ R\to\infty,
	\end{split}
\end{equation}
where $C>0$ is independent of $R>0$.
Therefore, we obtain that
\begin{align*}
	&E_\alpha(\lambda_1+\lambda_2)-E_\alpha(\lambda_1)-E^\infty_\alpha(\lambda_2)\\
	\leq&
	-\frac{\alpha^{2p-2}}{p}\int_{\R^3}\Big(\big(\rho_{\gamma_1}
	+\rho_{\gamma'_2}\big)^p-\rho_{\gamma_1}^p
	-\rho_{\gamma'_2}^p\Big)dx-CR^{-1}+O(a_R^2)\\
	\leq&-CR^{-1}+O(a_R^2)\ \ \ \mathrm{as}\ \ R\to\infty.
\end{align*}
where $C>0$ is as in \eqref{2.42}. On the other hand,  one can  check  from \eqref{11} and \eqref{2.7} that
$$
a_R=\max\limits_{i,j}\big|(u_i,v_j^R)\big|=o(R^{-\infty})\ \ \ \mathrm{as}\ \ R\to\infty,
$$
where $f(R):=o(R^{-\infty})$  means that  $ \lim\limits_{R\rightarrow\infty}f(R)R^s=0$ for any $s>0$.
As a consequence, we derive that
\begin{equation*}\label{str}
	E_\alpha(\lambda_1+\lambda_2)-E_\alpha(\lambda_1)-E^\infty_\alpha(\lambda_2)<-\frac{C}{2}R^{-1}<0,
\end{equation*}
if $R>0$ is sufficiently large. This therefore completes the proof of Lemma  \ref{lem2.4}.\qed

 \vspace {.5cm}

\noindent 
{\bf Data availability} Data sharing is not applicable to this article as no new data were created or analyzed in
this study.

 \vspace {.5cm}\noindent 
{\bf Conflict of interest} On behalf of all authors, the corresponding author states that there is no conflict of interest.


\begin{thebibliography}{99}
		
		
		
		
		
		
		
		
		
		
\bibitem{Cazenave} T. Cazenave,  Semilinear Schr\"{o}dinger Equations, Courant Lecture Notes in Mathematics Vol. {\bf10}, Courant Institute of Mathematical Science, New York, 2003.
		
\bibitem{critical} B. Chen, Y. J. Guo and S. Zhang, {\em Ground states of fermionic nonlinear Schr\"{o}dinger systems with Coulomb potential II: the $L^2$-critical case}, submitted (2024), 39 pages.
		
\bibitem{ims} H. L. Cycon, R. G. Froese, W. Kirsch and B. Simon,  Schr\"{o}dinger Operators with Application to Quantum Mechanics and Global Geometry, Texts and Monographs in Physics, Springer Study Edition, Springer-Verlag, Berlin, 1987.
		

\bibitem{l2} G. Fibich, The nonlinear Schr\"{o}dinger equation: singular solutions and optical collapse, Springer, 2015.

\bibitem{ii} R. L. Frank, D. Gontier and M. Lewin, {\em The nonlinear Schr\"{o}dinger equation for orthonormal functions II: application to Lieb-Thirring inequalities}, Comm. Math. Phys. {\bf 384} (2021), no. 3, 1783--1828.
		
		

		
\bibitem{elli} D. Gilbarg and N. S. Trudinger, Elliptic Partial Differential Equations, 2nd. Belin: Springer, 1997.
		
\bibitem{i} D. Gontier, M. Lewin and F. Q. Nazar, \emph{The nonlinear Schr$\ddot{o}$dinger equation for orthonormal functions: existence of ground states}, Arch. Ration. Mech. Anal. {\bf240} (2021), 1203--1254.
		
		
		
		
		
		
\bibitem{hq} Q. Han and F. H. Lin, Elliptic Partial Differential Equations, 2nd ed., Courant Lecture Notes in Mathematics Vol. {\bf 1}, Courant Institute of Mathematical Science/AMS, New York, 2011.
		
\bibitem{expo} P. D. Hislop, \emph{Exponential decay of two-body eigenfunctions: a review}, Proceedings of the Symposium on Mathematical Physics and Quantum Field Theory (Berkeley, CA, 1999), 265--288, Electron. J. Differ. Equ. Conf. 4, Southwest Texas State Univ. San Marcos, TX, 2000.
		

\bibitem{ho} M. Hoffmann-Ostenhof and T. Hoffmann-Ostenhof,  \emph{Schr\"{o}dinger inequalities and asymptotic behavior of the electron density of atoms and molecules}, Phys. Rev. A, {\bf 16} (1977), 1782--1785.
		
		
\bibitem{geomrtric} M. Lewin,  \emph{Geometric methods for nonlinear many-body quantum systems}, J. Funct. Anal. {\bf 260} (2011), 3535--3595.
		
		
		
		
\bibitem{analysis} E. H. Lieb and M. Loss,   Analysis, Graduate Studies in Mathematics Vol. 14, 2nd ed, American Mathematical Society, Providence, RI, 2001.
		
\bibitem{L3} E. H. Lieb and R. Seiringer,  The Stability of Matter in Quantum Mechanics, Cambridge University Press, 2010.
		
\bibitem{1974} E. H. Lieb and B. Simon, {\em On solutions to the Hartree-Fock problem for atoms and molecules}, J. Chem. Phys. {\bf 61} (1974),  735--736.
		
\bibitem{1977} E. H. Lieb and B. Simon, {\em The Hartree-Fock theory for Coulomb systems}, Commun. Math. Phys. {\bf 53} (1977), 185--194.
		
		
\bibitem{L2} E. H. Lieb and W. E. Thirring, \emph{Bound for kinetic energy of fermions which proves the stability of matter}, Phys. Rev. Lett. {\bf 35} (1975), 687--689.
		
		
\bibitem{concen} P. L. Lions,  \emph{The concentration-compactness principle in the calculus of variations. The locally compact case. I}, Ann. Inst. Henri Poincar\'{e}. Anal. Non Lin\'{e}aire {\bf1} (1984), 109--145.
		
\bibitem{concen2} P. L. Lions,  \emph{The concentration-compactness principle in the calculus of variations. The locally compact case. II}, Ann. Inst H. Poincar\'{e}. Anal. Non Lineaire  {\bf1} (1984), 223--283.

\bibitem{1987} P. L. Lions,  \emph{Solutions of Hartree-Fock equations for Coulomb systems}, Commun. Math. Phys. {\bf 109} (1987), 33--97.
		
\bibitem{m} M. Maeda,  \emph{On the symmetry of the ground states of nonlinear Schr\"{o}dinger equation with potential}, Adv. Nonlinear Stud. {\bf 10} (2010), 895--925.

\bibitem{begain} B. Marino, S. Enrico, Semilinear Elliptic Equations for Beginners, Springer, London, 2011.
		

\bibitem{Rabinowitz} P. H. Rabinowitz, \emph{On a class of nonlinear Schr\"{o}dinger equations}, Z. Angew. Math. Phys. {\bf 43} (1992), 270--291.
		
\bibitem{modern1} M. Reed and B. Simon, Methods of Modern Mathematical Physics I: Functional Analysis, Second edition, Academic Press, Inc. New York, 1980.		
		
\bibitem{modern} M. Reed and B. Simon,  Methods of Modern Mathematical Physics II: Fourier analysis, self-adjointness, Academic Press, New York-London, 1975.
		
\bibitem{modern4} M. Reed and B. Simon,  Methods of Modern Mathematical Physics IV: Analysis of operators, Academic Press, New York-London, 1978.
		
\bibitem{1926} E. Schr\"{o}dinger, {\em Quantisierung als Eigenwertproblem}, Ann. Phys. {\bf 81} (1926),  220--250.
		
		
\bibitem{minmax} M. Willem,  Minimax Theorems, Progress in Nonlinear Differential Equations and Their Applications Vol. {\bf 24}, Birkh\"{a}user Boston, Inc. Boston, 1996.
		
		
		
		
		
\end{thebibliography}
\end{document}